\documentclass[journal]{IEEEtran}

\usepackage{subeqn}
\usepackage{graphicx}
\usepackage{float}
\usepackage{mathrsfs}
\usepackage{amsfonts}
\usepackage{cite}
\usepackage{color}
\usepackage{subfigure}
\usepackage{booktabs}
\usepackage{multirow}

\usepackage{algorithm}
\usepackage{algpseudocode}
\usepackage{graphics}

\ifCLASSINFOpdf
\else
\fi
\begin{document}

\title{Joint Channel Estimation and Channel Decoding in Physical-Layer Network Coding Systems: An EM-BP Factor Graph Framework}
\author{Taotao Wang, \emph{Student Member, IEEE,} and Soung Chang Liew, \emph{Fellow, IEEE}
\thanks{The authors are with the Department of Information Engineering, The Chinese University of Hong Kong. Email: \{wtt011, soung\}@ie.cuhk.edu.hk   }
\thanks{Part of this work was presented at the IEEE International Conference on Acoustics, Speech, and Signal Processing (ICASSP), Vancouver, Canada, May, 2013.}
}
\markboth{Joint Channel Estimation and Channel Decoding in Physical-Layer Network Coding Systems: An EM-BP Factor Graph Framework}%
{Shell \MakeLowercase{\textit{et al.}}: Bare Demo of IEEEtran.cls
for Journals}
\maketitle
\vspace{-0.5in}

\begin{abstract}
This paper addresses the problem of joint channel estimation and channel decoding in physical-layer network coding (PNC) systems. In PNC, multiple users transmit to a relay simultaneously. PNC channel decoding is different from conventional multi-user channel decoding: specifically, the PNC relay aims to decode a network-coded message rather than the individual messages of the users. Although prior work has shown that PNC can significantly improve the throughput of a relay network, the improvement is predicated on the availability of accurate channel estimates. Channel estimation in PNC, however, can be particularly challenging because of 1) the overlapped signals of multiple users; 2) the correlations among data symbols induced by channel coding; and 3) time-varying channels. We combine the expectation-maximization (EM) algorithm and belief propagation (BP) algorithm on a unified factor-graph framework to tackle these challenges. In this framework, channel estimation is performed by an EM subgraph, and channel decoding is performed by a BP subgraph that models a virtual encoder matched to the target of PNC channel decoding. Iterative message passing between these two subgraphs allow the optimal solutions for both to be approached progressively. We present extensive simulation results demonstrating the superiority of our PNC receivers over other PNC receivers.
\end{abstract}


\IEEEpeerreviewmaketitle

\section{Introduction}
\label{sec:intro}


Recently, the research community has shown growing interest in a simple relay network in which two terminal nodes communicate via a relay. This network is referred to as the two-way relay channel (TWRC). Much of the interest in TWRC is on the exploitation of physical-layer network coding (PNC) \cite{zhang2006hot, liew2011physical} to boost its throughput.

Ref. \cite{zhang2006hot} showed that PNC could increase TWRC throughput by 100\% compared with traditional relaying \cite{liew2011physical}. In TWRC operated with PNC, the two terminal nodes first transmit their messages simultaneously to the relay. The relay then maps the overlapped signals to a network-coded message (e.g., bit-wise XOR of the messages of the terminal nodes) and broadcasts the network-coded message to the two terminal nodes. Each terminal node then extracts the message of the other terminal node by subtracting its own message from the network-coded message. Thus, the two terminal nodes exchange one message with each other in two time slots. With traditional relaying, four time slots are needed \cite{zhang2006hot}.

This throughput advantage of PNC, however, is predicated on the accurate estimate of the channels between the terminal nodes and the relay. For optimality, it is desirable to obtain the maximum \emph{a posteriori} probability (MAP) channel estimates. This is, however, a particularly challenging task for PNC. The involved issues addressed by this paper are as follows:

\begin{itemize}
 \item  	For reliable communication, we consider channel-coded PNC \cite{zhang2009channel, liew2011physical }. Specifically, the source messages of the two terminal nodes are channel-coded into channel-coded messages before transmission. The signals received by the relay contain the overlapped channel-coded messages as well as overlapped preambles (training symbols) and pilots.
 \item  	PNC channel decoding is different from conventional multiuser channel decoding. The goal of the PNC channel decoding at the relay is to obtain a network-coded message rather than the two individual source messages \cite{zhang2009channel, liew2011physical }. In other words the relay aims not to decode the two source messages, but to decode a network-coding function of the two source messages (in this paper, we assume the network-coding function is the bit-wise XOR of the two source messages).
 \item  	For successful decoding, accurate channel estimates are needed. For optimality, it is desirable to estimate the channels using not just the preambles and pilots, but also the data in the signals. This is because the data portion also contains useful information related to the channels.
 \item  	We are interested in time-varying channels in which the channel gains vary from symbol to symbol within a packet.
\end{itemize}
Overall, performing channel estimation and PNC channel decoding when (i) the signals are overlapping; (ii) the data symbols are correlated due to channel coding; and (iii) the channels are time-varying, is a particularly challenging task.

To tackle this challenge, this paper proposes and investigates a joint channel estimation and channel decoding framework.  We argue that directly trying to solve the MAP channel estimation problem and the channel decoding problem in a separate manner is not viable; a solution is found in a combined use of expectation-maximization (EM) algorithm and belief propagation (BP) algorithm that solves the two problems jointly in an iterative manner.

We implement the EM-BP computation as a message passing algorithm on a factor graph \cite{kschischang2001factor, loeliger2007factor}, in which the component for channel estimation (implemented by EM) and the component for PNC channel decoding (implemented by BP) are interconnected.  Through iterative message passing between the EM and BP components and iterative message passing between elements within the BP channel decoding component, the results of EM channel estimation and BP channel decoding improve progressively toward optimality.

Overall, there are three major contributions to this work:
\begin{enumerate}
\item This is the first work that applies the EM algorithm for joint channel estimation and channel decoding in PNC systems.  Notably, our algorithmic framework includes a schema to deal with time-varying channels. To reduce computation complexity, we further extend our framework by replacing EM with its variant named space alternating generalized expectation-maximization.
 \item We outline a factor graph framework for iterative message passing algorithm based on the foundations of EM/SAGE and BP. This is the first time EM/SAGE-BP computation for a channel-coded communication system is fully implemented as a message passing algorithm on a factor graph (prior work either did not consider channel coding or did not use the factor graph schema). In particular, we explicitly establish this EM/SAGE-BP factor graph framework from a rigorous theoretical foundation. We remark that although our focus here is on PNC systems, this framework is also applicable to the conventional single-user system and multi-user system.
 \item Through extensive computer simulations, we investigate the performance of our EM/SAGE-BP PNC receivers and compare them with other existing receivers in the literatures.  The simulation results demonstrate the superiority of our receivers over other receivers, confirming the theoretical optimality of EM/SAGE-BP PNC.
\end{enumerate}

\subsection{Related Works}

\noindent \textbf{Theory of BP and Its Application in PNC}:

BP, also known as the sum-product message passing algorithm, is a general algorithmic inference method for graph models \cite{kschischang2001factor, yedidia2003understanding}. It has found great success in the decoding of powerful channel codes (e.g., Turbo codes and LDPC codes \cite{mceliece1998turbo}) in point-to-point communications. BP was first applied to PNC channel decoding in \cite{zhang2009channel}, which puts forth the concept of `virtual encoding'. The concept was further generalized in \cite{wubben2010generalized, lu2012asynchronous}. Refs. \cite{zhang2009channel, wubben2010generalized, lu2012asynchronous} were followed by many other papers on channel-coded PNC \cite{liew2011physical}. The prior studies of PNC channel decoding mostly assume that the channels are perfectly known. In practice, these channels have to be estimated. PNC systems are multi-user systems in which there are multiple channel parameters, whose estimation is particularly challenging.

It is desired that we could jointly solve channel estimation and channel decoding by BP. Recently, \cite{lehmann2013joint} proposed a BP method for joint channel estimation and channel decoding in PNC systems. The direct application of BP to channel estimation, however, requires summations (integrations) over continuous channel variables, which are computationally intensive. To reduce computation complexity, a moment matching (MM) technique is used in \cite{lehmann2013joint}. Specifically, the original Gaussian-mixture messages in the BP algorithm are approximated by Gaussian messages that have the same first and second moments. However, as a consequence of the approximation by MM, the optimality cannot be guaranteed. By contrast, in this paper, instead of using a pure BP method for both channel estimation and channel decoding, we use an EM-BP method that obviates the need for computation-intensive integration without sacrificing optimality.

\noindent \textbf{Theory of EM and Its Application in Single-user Systems}:

Ref. \cite{dempster1977EM} first proposed EM as a general iterative algorithm for finding the maximum likelihood (ML) estimates of parameters in statistical model with hidden variables that cannot be observed directly. A small extension allows the finding of the MAP estimates also \cite{gupta2011theory}. A variant of EM named SAGE was proposed in \cite{sage1994} to increase the convergence rate. Recently, \cite{dauwels2005expectation, dauwels2009expectation} proposed a framework to implement EM computation as a message passing algorithm on a factor graph. Application for communications systems (and in particular, channel-coded communications systems) was not its main target. It is not clear that the factor graph representation in \cite{dauwels2005expectation, dauwels2009expectation} is applicable to the communication problem of interest to us here. With respect to our contribution 2) listed above, in this paper we fill in the missing details and show specifically how a factor graph representation can be used in our problem.  In addition, we extend the framework of EM message passing over factor graphs to SAGE  message passing.

Papers \cite{dempster1977EM, gupta2011theory, sage1994, dauwels2005expectation, dauwels2009expectation} concern  EM/SAGE in general. There have also been many investigations on the application of EM/SAGE in communication systems  specifically. Refs. \cite{georghiades1997sequence, chiavaccini2001map, cozzo2003joint, bcjrem, torrieri2006iterative, herzet2007theoretical, guo2011based} for example, applied EM to the problem of joint channel estimation and detection/decoding in single-user systems. Refs. \cite{ylioinas2009iterative, panayirci2010joint} applied SAGE to the same problem.  The work \cite{guo2011based} directly applied the results in \cite{dauwels2005expectation, dauwels2009expectation} to implement EM message passing for joint channel estimation and detection in a single-user system. However, the incorporation of channel codes by \cite{guo2011based} is heuristic and is not in accordance with the rigorous theoretical derivation in Section III of our paper here.

\noindent \textbf{EM Application in Multi-user Systems}:

Refs. \cite{fawer1995multiuser, kocian2003based, wu2008iterative, assra2010based} applied EM to joint channel estimation and multi-user detection in CDMA systems. Channel coding was not considered. Ref. \cite{kopbayashi2001successive} incorporated channel coding. However, the proposed scheme performs serial interference cancelation (SIC) and tries to decode the individual messages of different users using separate channel decoders. Decoding individual messages is an overkill for PNC systems and may lead to suboptimal performance (we will provide results showing this in Section V). The application of SAGE in multi-user systems for joint channel estimation and detection can be found in \cite{kocian2003based, kocian2007joint, pun2007iterative}, where \cite{kocian2003based, kocian2007joint} assume CDMA systems and \cite{pun2007iterative} assumes OFDMA systems.

Overall, there has been little multi-user EM work that incorporates channel coding. Furthermore, with respect to our contribution 1) listed above, it is worth re-emphasizing that PNC channel decoding \cite{zhang2009channel,  wubben2010generalized, lu2012asynchronous} is different from conventional multiuser channel decoding \cite{kopbayashi2001successive}, because PNC aims to channel-decode the overlapped received signals into a network-coded message \cite{zhang2009channel,  wubben2010generalized, lu2012asynchronous} rather than the individual messages of different users.

The rest of this paper is organized as follows. Section II describes our system model. In section III, we apply EM-BP to PNC and derive the update equations for EM-BP message passing. Section IV applies SAGE-BP to PNC. Section V presents the simulation results. Section VI concludes this paper.

\textbf{Notations}: We denote matrices by bold capital letters, vectors by bold small letters, and scalars by regular letters throughout this paper. All vectors are column vectors. ${{\bf{I}}_r}$ denotes the $r \times r$ identity matrix and  ${{\bf{0}}_r}$ denotes the $r \times r$ matrix with all zero elements. ${{\bf{A}}^{\rm{T}}}$, ${{\bf{A}}^{\rm{H}}}$, ${{\bf{A}}^{ - 1}}$ and $\det \left( {\bf{A}} \right)$ denote the transpose, the conjugate transpose, the inverse and the determinant of ${\bf{A}}$, respectively. ${\cal C}{\cal N}\left( {{\bf{x}}:{\bf{m}},{\bf{K}}} \right) \buildrel \Delta \over = \frac{1}{{{\pi ^r}\det \left( {\bf{K}} \right)}}\exp \left[ { - \left( {{\bf{x}} - {\bf{m}}} \right)^{\rm{H}}{{\bf{K}}^{ - 1}}{{\left( {{\bf{x}} - {\bf{m}}} \right)}}} \right]$
denotes the probability density function (PDF) of an $r$-dimension complex Gaussian random variable ${\bf{x}}$  with mean vector ${\bf{m}}$  and covariance matrix ${\bf{K}}$ . $\left\| {\bf{x}} \right\|$ is the Euclidean norm of a vector ${\bf{x}}$ . ${\left(  \cdot  \right)^ * }$ denotes the conjugate of a complex number. $\left| C \right|$ is the cardinality of a set $C$.  ${x_{i:j}}$ is a set containing the elements in a sequence $x$  indexed from $i$  to $j$, inclusively. Finally,  $\otimes$ denotes the Kronecker product, and $\oplus $
 denotes the exclusive-or operation.

\section{System Model}

We consider a two-phase PNC transmission scheme for TWRC consisting of an uplink phase and a downlink phase. In the uplink phase, two terminal nodes A and B transmit packets to a relay node R simultaneously. From the overlapped signals received from A and B, R constructs a network-coded packet and broadcasts it to A and B in the downlink phase. From the network-coded packet, A(B) then recovers the packet of B(A) using its self information \cite{liew2011physical}.

This paper focuses on the uplink phase; the problem of reliably transmitting the network-coded packet in the downlink phase is similar to that in a conventional point-to-point link. We assume A and B have one transmit antenna, and R has one receive antenna. In the uplink phase, the received signal at R in the $i^{th}$ symbol duration can be expressed as
\begin{equation}
\begin{array}{l}
 {y_i} = {h_{A,i}}{x_{A,i}} + {h_{B,i}}{x_{B,i}} + {n_{R,i}} = {\bf{h}}_i^{\rm{T}}{{\bf{x}}_i} + {n_{R,i}} \\
 \end{array}
\end{equation}
where ${x_{A,i}}\left( {{x_{B,i}}} \right)$ is the ${i^{th}}$ transmitted symbol of node A(B); ${{h}}_i^{\rm{A}}\left( {{{h}}_i^{\rm{B}}} \right)$ is the ${i^{th}}$  fading coefficient of the channel between  A(B) and R; ${{n}_{R,i}}$ is the complex white Gaussian noise sample with zero mean and variance $N_{0}$; ${{\bf{h}}_i} \buildrel \Delta \over = {[h_i^{\rm{A}},h_i^{\rm{B}}]^{\rm{T}}}$; and ${{\bf{x}}_i} \buildrel \Delta \over = {\left[ {{x_{A,i}},{x_{B,i}}} \right]^{\rm{T}}}$. A block diagram of the system model is shown in Fig. 1, where $\left\{ {{s_{A,j}}} \right\}$ and  $\left\{ {{s_{B,j}}} \right\}$ are the source information bits from nodes A and B. The transmitted symbols  $\left\{ {{x_{A,i}}} \right\}$ and  $\left\{ {{x_{B,i}}} \right\}$ are generated after channel encoding, interleaving, constellation mapping and pilot insertions at the transmitters. In this work, we assume that A and B use the same channel encoder\footnote{Discussion on how to deal with two nodes with different channel encoders can be found in Section IV.} (the valid set of codewords is denoted by $C$) and the same interleaver when mapping their source bits $\left\{ {{s_{A,j}}} \right\}$ and  $\left\{ {{s_{B,j}}} \right\}$  to transmitted symbols. Pilot symbols are inserted periodically among coded data symbols. The assumed frame structure is shown in Fig. 2 where $P$  and $D$ represent the pilot symbols and coded data symbols, respectively.  Each frame consists of $l$  data symbols, divided into ${l \mathord{\left/
 {\vphantom {l \Delta }} \right.
 \kern-\nulldelimiterspace} \Delta }$ blocks.  Each block has $ \Delta$ data symbols and two pilot symbols. We refer to $ \Delta$  as the pilot interval.\footnote{Simulation results on the impact of the pilot interval on system performance can be found in Section V.} The total frame length is $L = l + 2\left( {{l \mathord{\left/
 {\vphantom {l \Delta}} \right.
 \kern-\nulldelimiterspace} \Delta}} \right)$ symbols.

\begin{figure}[!t]
\centering
\includegraphics[width=3.5in]{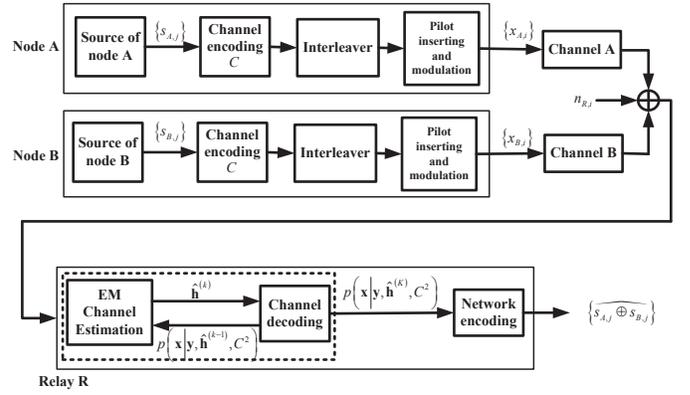}
\caption{The system model of uplink phase in TWRC.} \label{system_model}
\end{figure}

\begin{figure}[!t]
\centering
\includegraphics[width=3.5in]{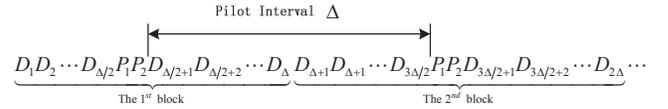}
\caption{The frame structure employed by the two user nodes.} \label{frameformat}
\end{figure}

We assume time-varying Rayleigh fading channels. The time-varying channel gains $\left\{ {{{{h}}_{A,i}}} \right\}$ and  $\left\{ {{{{h}}_{B,i}}} \right\}$ are modeled as two independent first-order Gauss-Markov processes \cite{tan2000first}:
\begin{equation}
\begin{array}{l}
 {{{h}}_{A,i}} = \alpha {{{h}}_{A,i - 1}} + \sqrt {1 - {\alpha ^2}} {{{z}}_{A,i}} \\
 {{{h}}_{B,i}} = \alpha {{{h}}_{B,i - 1}} + \sqrt {1 - {\alpha ^2}} {{{z}}_{B,i}} \\
 \end{array}
\end{equation}
where $\left\{ {{{{z}}_{A,i}}} \right\}$ and  $\left\{ {{{{z}}_{B,i}}} \right\}$ are independent white complex Gaussian processes with zero mean and variances $\sigma _{\rm{A}}^2$ and $\sigma _{\rm{B}}^2$ for all $i$. The parameter $\alpha$ is a correlation coefficient used for modeling how fast the channel varies with time (i.e., it is related to the channel coherence time) \cite{tan2000first}. The distributions of ${{{h}}_{A,i}}$ and  ${{{h}}_{B,i}}$  are zero mean complex Gaussian with variances $\sigma _A^2$ and $\sigma _B^2$, respectively. Therefore, the amplitude of every element of them is Rayleigh distributed, and the phase is uniformly distributed.


\section{Application of EM-BP to PNC}

\subsection{Objectives of EM PNC Receiver}

Let ${\bf{h}} \buildrel \Delta \over = \left\{ {{{\bf{h}}_i}} \right\}$ be the set containing channel gains of all time. Similarly, let ${\bf{x}}$ be the set of all transmitted symbols $\left\{ {{{\bf{x}}_i}} \right\}$, and ${\bf{y}}$  is the set of all received signals $\left\{ {{{{y}}_i}} \right\}$.  To the relay, both ${\bf{h}}$   and  ${\bf{x}}$ are unknowns to be estimated and decoded.

In a conventional receiver, ${\bf{h}}$ is first estimated, followed by the decoding of codewords ${\bf{x}}$.  Pilots, corresponding to known ${{\bf{x}}_i}$ at specific positions $i$, are used for the estimation of ${\bf{h}}$.  After that, ${{\bf{h}}_i}$ for the positions occupied by data are estimated by interpolation. This estimate of ${\bf{h}}$  is then substituted into (1) for the decoding of the data ${{\bf{x}}_i}$. This estimate of ${\bf{h}}$ makes use of only the pilot parts, and does not exploit useful information contained in the data part of ${\bf{x}}$.

In our PNC receiver design, we wish to make use of the pilots as well as the data parts of ${\bf{x}}$ in the channel estimation process. In particular, observations ${\bf{y}}$ at all positions $i$ and the knowledge of the used channel code will be employed to estimate ${\bf{h}}$.  A possible strategy for our channel estimation and channel decoding problem in PNC receiver is as follows:

\noindent \textbf{Step 1 (channel estimation)}:  Find MAP estimate ${\widehat{\bf{h}}_{{\rm{MAP}}}} = \arg \mathop {\max }\limits_{\bf{h}} \left\{ {\log p\left( {{\bf{h}}\left| {{\bf{y}},{C^2}} \right.} \right)} \right\} = \arg \mathop {\max }\limits_{\bf{h}} \left\{ {\log \sum\limits_{\bf{x}} {p\left( {{\bf{x}},{\bf{h}}\left| {{\bf{y}},{C^2}} \right.} \right)} } \right\}$

\noindent  \textbf{Step 2 (channel decoding)}: Given ${\widehat{\bf{h}}_{{\rm{MAP}}}}$, find $p\left( {{\bf{x}}\left| {{{\widehat{\bf{h}}}_{{\rm{MAP}}}},{\bf{y}},{C^2}} \right.} \right)$

\noindent  \textbf{Step 3 (network coding)}: Compute the network-coded source message $\left\{ {\widehat{{s_{A,j}} \oplus {s_{B,j}}}} \right\}$  based on the channel decoding output from Step 2 \cite{liew2011physical}.

\noindent  This is the PNC receiver with optimal channel estimation. Some remarks on the receiver are as follows:

\begin{enumerate}
 \item A subtlety here in Step 2 for the PNC system is that ${C^2}$  is the code constraint by the `virtual channel encoder' which takes the original information source symbols from nodes A and B $\left\{ {{s_{A,j}},{s_{B,j}}} \right\}$  as inputs, and outputs $\left\{ {{{\bf{x}}_i}} \right\}$ as coded symbols  (see \cite{liew2011physical} for details).

 \item If the channel coefficients were perfectly known (as assumed in previous works \cite{zhang2009channel, wubben2010generalized, lu2012asynchronous}), then Step 1 would  not be needed. Step 2 and Step 3 then form the so-called Channel-decoding-Network-Coding (CNC) process, an essence of channel-coded PNC systems \cite{liew2011physical, zhang2009channel}. Compared to conventional channel decoding, the goal of CNC is not to decode the individual source messages of A and B, but to decode a network-coded message that mixes the two source messages (we refer the interested readers to  \cite{liew2011physical} and references therein for details on CNC). If the MAP estimation in Step 1 could be achieved, then Step 2 and Step 3 could be implemented using the conventional CNC methods, substituting ${\widehat{\bf{h}}_{{\rm{MAP}}}}$ as the channel coefficients \cite{liew2011physical, zhang2009channel}.  Unfortunately, this is not viable because the exact MAP estimate of ${\bf{h}}$  is difficult due to the complexity of the computation of $\sum\nolimits_{\bf{x}} {p\left( {{\bf{h}},{\bf{x}}\left| {\bf{y}} \right.} \right)}$. A difficulty, for example, is that the symbols in ${\bf{x}}$ are correlated due to channel coding; in addition, signals of the two terminal nodes are overlapped in $\bf{y}$.

\end{enumerate}

To facilitate the design of PNC receiver, this paper makes use of the EM algorithm for channel estimation. Specifically, EM tries to find ${\widehat{\bf{h}}_{{\rm{MAP}}}}$ iteratively rather than attacking the problem directly.  The objective of EM is still to obtain the MAP estimate of  ${\bf{h}}$ as in Step 1. However, EM combines Step 1 and Step 2 in an iterative manner to refine the estimate of ${\bf{h}}$ and the decoding of the network-coded message.  In the following, we first describe the procedure of the EM computation and then present its implementation as a message passing algorithm on a factor graph.

In the terminology of EM,  ${\mathbf{y}}$  is the observed data, ${\mathbf{x}}$  is the hidden data, the pair $\left( {{\mathbf{x}},{\mathbf{y}}} \right)$ is the complete data, and ${\mathbf{h}}$  is the unknown parameter. The  $k^{th}$  iteration of EM consists of an E-step (expectation) and an M-step (maximization) as follows \cite{gupta2011theory}:

\noindent 	\textbf{E-step}: Given the previous estimate ${\widehat{\mathbf{h}}^{\left( {k - 1} \right)}}$, compute the conditional expectation
 \begin{equation}\label{E-step}
 Q\left( {{\mathbf{h}}\left| {{{\widehat{\mathbf{h}}}^{\left( {k - 1} \right)}}} \right.} \right) = \sum\limits_{\mathbf{x}} {p\left( {{\mathbf{x}}\left| {{\mathbf{y}},{{\widehat{\mathbf{h}}}^{\left( {k - 1} \right)}},{C^2}} \right.} \right)} \log p\left( {{\mathbf{y}},{\mathbf{x}}|{\mathbf{h}},{C^2}} \right)
 \end{equation}

\noindent   	\textbf{M-step}: Then, compute ${\widehat{\bf{h}}^{\left( k \right)}}$ by
 \begin{equation}\label{M-step}
 {\widehat{\bf{h}}^{\left( k \right)}} = \arg \mathop {\max }\limits_{\bf{h}} \left[ {Q\left( {{\bf{h}}\left| {{{\widehat{\bf{h}}}^{\left( {k - 1} \right)}}} \right.} \right) + \log p({\bf{h}})} \right]
 \end{equation}
The E-step in (3) can be broken down as follows. First, compute $p\left( {{\bf{x}}\left| {{\bf{y}},{{\widehat{\bf{h}}}^{\left( {k - 1} \right)}},{C^2}} \right.} \right)$ from ${\bf{y}}$
and ${\widehat{\bf{h}}^{\left( {k - 1} \right)}}$. This computation of $p\left( {{\bf{x}}\left| {{\bf{y}},{{\widehat{\bf{h}}}^{\left( {k - 1} \right)}},{C^2}} \right.} \right)$  is similar to Step 2 above, with ${\widehat{\bf{h}}^{\left( {k - 1} \right)}}$ replacing ${\widehat{\bf{h}}_{{\rm{MAP}}}}$. If the algorithm were to stop at iteration $k-1$ ,  we could simply go to Step 3 to obtain the network-coded message based on $p\left( {{\bf{x}}\left| {{\bf{y}},{{\widehat{\bf{h}}}^{\left( {k - 1} \right)}},{C^2}} \right.} \right)$. Otherwise, the E-step continues and uses $p\left( {{\bf{x}}\left| {{\bf{y}},{{\widehat{\bf{h}}}^{\left( {k - 1} \right)}},{C^2}} \right.} \right)$  to compute the $Q\left( {{\bf{h}}\left| {{{\widehat{\bf{h}}}^{\left( {k - 1} \right)}}} \right.} \right)$ in (3). After that, the M-step finds a new estimate of ${\bf{h}}$ as in (4).  In this way, the computation procedure jointly estimates the channels ${\bf{h}}$ and decodes the data ${\bf{x}}$, refining the solutions from one iteration to the next.  This process is shown in the lower half of Fig. 1, which depicts our PNC receiver.

Let us now re-examine (3) and (4) to  explain why they work. Combining the E-step and M-step by substituting (3) into (4) and after some manipulations, we find that the EM algorithm is actually a fixed-point iteration algorithm as follows:	
\begin{equation}
\begin{array}{l}
 {\widehat{\bf{h}}^{\left( k \right)}} = \arg \mathop {\max }\limits_{\bf{h}} \left[ {\sum\limits_{\bf{x}} {p\left( {{\bf{x}}\left| {{\bf{y}},{{\widehat{\bf{h}}}^{\left( {k - 1} \right)}},{C^2}} \right.} \right)} \log p\left( {{\bf{y}},{\bf{x}}|{\bf{h}},{C^2}} \right)} \right. \\
 \;\;\;\;\;\;\;\;\; \;\;\;\;\;\;\;\;\; \;\;\;\;\;\;\;\;\; \;\;\;\;\;\;\;\;\; \;\;\;\;\;\;\;\;\; \;\;\;\;\;\;\;\;\;  \;\;\;\;\;\;\;\;\; \;\;\;\;\;\;\;\;\;    \left. { + \log p({\bf{h}})} \right] \\
 =\arg \mathop {\max }\limits_{\bf{h}} \left[ { - {D_{KL}}\left( {p\left( {{\bf{x}}\left| {{\bf{y}},{{\widehat{\bf{h}}}^{\left( {k - 1} \right)}},{C^2}} \right.} \right)||p\left( {{\bf{x}}\left| {{\bf{y}},{\bf{h}},{C^2}} \right.} \right)} \right)} \right. \\
  \;\;\;\;\;\;\;\;\; \;\;\;\;\;\;\;\;\; \;\;\;\;\;\;\;\;\; \;\;\;\;\;\;\;\;\; \;\;\;\;\;\;\;\;\; \;\;\;\;\;\;\;\;\;  \;\;\;\;\;\; \left. { + \log p\left( {{\bf{h}}\left| {{\bf{y}},{C^2}} \right.} \right)} \right] \\
 \end{array}
\end{equation}
where  ${D_{KL}}$ is the Kullback-Leibler (K-L) divergence. It is known that miminizing K-L divergence ${D_{KL}}$ corresponds to minimizing the difference between two distributions \cite{cover2012elements}. The last line of (5), however, is not exactly minimizing ${D_{KL}}$; besides ${D_{KL}}$ , there is an additional term $\log p\left( {{\bf{h}}\left| {{\bf{y}},{C^2}} \right.} \right)$  in the function be optimized. Another way to look at (5) is as follows. Recall that finding $\arg {\max _{\bf{h}}}\log p\left( {{\bf{h}}\left| {\bf{y}} \right.,{C^2}} \right)$ is the original objective of Step 1. In (5), this objective is modified by the additional term ${D_{KL}}$. The appendix shows that the fixed-point EM iteration in (5) is still an attempt to find $\arg {\max _{\bf{h}}}\log p\left( {{\bf{h}}\left| {\bf{y}} \right.,{C^2}} \right)$  despite the additional ${D_{KL}}$ term. Indeed, finding $\arg {\max _{\bf{h}}}\log p\left( {{\bf{h}}\left| {\bf{y}} \right.,{C^2}} \right)$  is facilitated by adding ${D_{KL}}$, which goes to zero when the algorithm converges. The solution of EM is at least a local optimal $\log p({\widehat{\bf{h}}_{{\rm{MAP}}}}\left| {{\bf{y}},{C^2}} \right.)$ although it may not be the global optimal. A good initial point will often lead to a global optimal \cite{wu1983convergence}.

\subsection{Implementation of EM-BP PNC on Factor Graph}

Although EM can iteratively find the MAP estimate, the computations of the E-step (3) and M-step (4) are still non-trivial. We next establish a framework that implement the EM computation as a message passing algorithm on a factor graph, where we can rigorously combine EM message passing for channel estimation with BP message passing for channel decoding. This framework allows us to more clearly see the intricacies of the EM computation, pointing to a systematic and practical way to implement it.

Refs \cite{dauwels2005expectation, dauwels2009expectation} showed how to transform a generic EM computation to a factor graph implementation. It is not clear, however, that the assumptions in \cite{dauwels2005expectation, dauwels2009expectation} on the functional forms of the parameter and variable probabilities are valid for our specific problem.  Here, we give a theoretical derivation tailored to channel-coded communication systems.

A key to factor graph implementation is to factorize $p\left( {{\bf{y}},{\bf{x}}|{\bf{h}},{C^2}} \right)$ in (\ref{E-step}) and $p({\bf{h}})$ in (\ref{M-step}).  For $p\left( {{\bf{y}},{\bf{x}}|{\bf{h}},{C^2}} \right)$, we write
\begin{equation}\label{f1}
\begin{array}{l}
 p\left( {{\bf{y}},{\bf{x}}|{\bf{h}},{C^2}} \right) = p\left( {{\bf{y}}\left| {{\bf{x}},{\bf{h}}} \right.} \right)p\left( {{\bf{x}}\left| {{C^2}} \right.} \right)
= \frac{{{I_{{C^2}}}\left( {\bf{x}} \right)\prod\nolimits_i {p\left( {{{{y}}_i}\left| {{{\bf{x}}_i},{{\bf{h}}_i}} \right.} \right)} }}{{\left| {{C^2}} \right|}} \\
 \end{array}
\end{equation}
where ${I_{{C^2}}}\left( {\bf{x}} \right)$ is a indicator function defined as: ${I_{{C^2}}}\left( {\bf{x}} \right) = 1$ if ${\bf{x}} \in {C^2}$;  ${I_{{C^2}}}\left( {\bf{x}} \right) = 0$ if ${\bf{x}} \notin {C^2}$, where we have assumed all codewords are equally likely. Note that we have used (1) to justify the factorization of $p\left( {{\bf{y}}\left| {{\bf{x}},{\bf{h}}} \right.} \right)$ into  $\prod\nolimits_i {p\left( {{{{y}}_i}\left| {{{\bf{x}}_i},{{\bf{h}}_i}} \right.} \right)}$
in (\ref{f1}), where ${{{y}}_i}$ is independent of ${{{y}}_j}$  for all $i \ne j$ given ${{\bf{x}}}$ and ${{\bf{h}}}$ thanks to their independent noise components. Substituting (\ref{f1}) into the $Q$  function in (3) and dropping the term $ - \log \left| {{C^2}} \right|$, which is independent of ${\bf{h}}$ and therefore does not matter as far as the M-step is concerned, we have
\begin{equation}
\begin{array}{l}
 Q\left( {{\bf{h}}\left| {{{\widehat{\bf{h}}}^{\left( {k - 1} \right)}}} \right.} \right)   \\ =
 \sum\limits_i {\sum\limits_{{{\bf{x}}_i}} {\log p\left( {{{{y}}_i}\left| {{{\bf{x}}_i},{{\bf{h}}_i}} \right.} \right)} \underbrace {\sum\limits_{{{\bf{x}}_{1:i - 1}},{{\bf{x}}_{i + 1:L}}} {p\left( {{\bf{x}}\left| {{\bf{y}},{{\widehat{\bf{h}}}^{\left( {k - 1} \right)}},{C^2}} \right.} \right)} }_{ \buildrel \Delta \over = p\left( {{{\bf{x}}_i}\left| {{\bf{y}},{{\widehat{\bf{h}}}^{\left( {k - 1} \right)}},{C^2}} \right.} \right)}}  \\
 \end{array}
\end{equation}
where $p\left( {{{\bf{x}}_i}\left| {{\bf{y}},{{\widehat{\bf{h}}}^{\left( {k - 1} \right)}},{C^2}} \right.} \right)$ is the \emph{a posteriori probability} (APP)\footnote{ If ${{\bf{x}}_i}$
is a vector of coded symbols, its APP is given by the soft channel decoder. If it is a vector of pilot symbols, this probability is either 0 or 1, given by the \emph{a priori} information available at the receiver about the pilots.} that can be computed using BP (sum-product) message passing algorithm for channel decoding on the factor graph  \cite{kschischang2001factor} of the given channel code ${C^2}$, with a fixed channel ${\widehat{\bf{h}}^{\left( {k - 1} \right)}}$.  We define the symbol-wise $Q$  function as
\begin{equation}
{Q_i}\left( {{{\bf{h}}_i}\left| {{{\widehat{\bf{h}}}^{\left( {k - 1} \right)}}} \right.} \right) \buildrel \Delta \over = \sum\limits_{{{\bf{x}}_i}} {\log p\left( {{{{y}}_i}\left| {{{\bf{x}}_i},{{\bf{h}}_i}} \right.} \right)p\left( {{{\bf{x}}_i}\left| {{\bf{y}},{{\widehat{\bf{h}}}^{\left( {k - 1} \right)}},{C^2}} \right.} \right)}
\end{equation}
With complex Gaussian white noise, the above $\log p\left( {{{{y}}_i}\left| {{{\bf{x}}_i},{{\bf{h}}_i}} \right.} \right)$ as a function of the variables ${{\bf{x}}_i}$ and  ${{\bf{h}}_i}$ can be obtained in closed form as ${{ - {{\left\| {{y_i} - {\bf{h}}_i^{\rm{T}}{{\bf{x}}_i}} \right\|}^2}} \mathord{\left/
 {\vphantom {{ - {{\left\| {{y_i} - {\bf{h}}_i^{\rm{T}}{{\bf{x}}_i}} \right\|}^2}} {N_{0}}}} \right.
 \kern-\nulldelimiterspace} {N_{0}}}{\rm{ }}$. We see that once $p\left( {{{\bf{x}}_i}\left| {{\bf{y}},{{\widehat{\bf{h}}}^{\left( {k - 1} \right)}},{C^2}} \right.} \right)$ is computed by BP channel decoding,  ${Q_i}\left( {{{\bf{h}}_i}\left| {{{\widehat{\bf{h}}}^{\left( {k - 1} \right)}}} \right.} \right)$ as a function of ${{\bf{h}}_i}$  can be obtained by the weighted sum of $p\left( {{{\bf{x}}_i}\left| {{\bf{y}},{{\widehat{\bf{h}}}^{\left( {k - 1} \right)}},{C^2}} \right.} \right)$
with weight ${{ - {{\left\| {{y_i} - {\bf{h}}_i^{\rm{T}}{{\bf{x}}_i}} \right\|}^2}} \mathord{\left/
 {\vphantom {{ - {{\left\| {{y_i} - {\bf{h}}_i^{\rm{T}}{{\bf{x}}_i}} \right\|}^2}} {N_{0}}}} \right.
 \kern-\nulldelimiterspace} {N_{0}}}{\rm{ }}$ over different possible values of ${{\bf{x}}_i}$. The overall $Q$  function is the sum of symbol-wise $Q$ functions:
\begin{equation}
Q\left( {{\bf{h}}\left| {{{\widehat{\bf{h}}}^{\left( {k - 1} \right)}}} \right.} \right) = \sum\nolimits_{i=1}^{L} {{Q_i}\left( {{{\bf{h}}_i}\left| {{{\widehat{\bf{h}}}^{\left( {k - 1} \right)}}} \right.} \right)}
\end{equation}
Using (9), the M-step in (4) is equivalent to
\begin{equation}
{\widehat{\bf{h}}^{\left( k \right)}} = \arg \mathop {\max }\limits_{\bf{h}} \left( {p\left( {\bf{h}} \right) \cdot \mathop \prod \nolimits_{i=1}^{L} {e^{{Q_i}\left( {{{\bf{h}}_i}\left| {{{\widehat{\bf{h}}}^{\left( {k - 1} \right)}}} \right.} \right)}}} \right){\rm{ }}
\end{equation}

To see what will happen in the M-step, let us employ the Gauss-Markov channel model (2) and factorize $p\left( {\bf{h}} \right)$ as
\begin{equation}
p\left( {\bf{h}} \right) = \prod\nolimits_{i = 1}^L {p\left( {{\bf{h}}_i \left| {{\bf{h}}_{i - 1} } \right.} \right)}
\end{equation}
where $p\left( {{\bf{h}}_1 \left| {{\bf{h}}_0 } \right.} \right) = p\left( {{\bf{h}}_1 } \right) = {\cal C}{\cal N}\left( {{\bf{h}}_1 :{\bf{0}},{\bf{Q}}} \right)$ with ${\bf{Q}} \buildrel \Delta \over = diag\left( {\left[ {\sigma _A^2,\sigma _B^2} \right]} \right)$ is the priori information of ${{\bf{h}}_1 }$, and $p\left( {{{\bf{h}}_i}\left| {{{\bf{h}}_{i - 1}}} \right.} \right) = {\cal C}{\cal N}\left( {{{\bf{h}}_i}:{{\bf{m}}_{{{\bf{h}}_i}\left| {{{\bf{h}}_{i - 1}}} \right.}},{{\bf{K}}_{{{\bf{h}}_i}\left| {{{\bf{h}}_{i - 1}}} \right.}}} \right)$ for $i \ge 2$ is the Markovian transition probability defined by (2). Specifically,  ${{\bf{m}}_{{{\bf{h}}_i}\left| {{{\bf{h}}_{i - 1}}} \right.}} = \alpha {{\bf{h}}_{i - 1}}$, ${{\bf{K}}_{{{\bf{h}}_i}\left| {{{\bf{h}}_{i - 1}}} \right.}} = \left( {1 - {\alpha ^2}} \right){\bf{Q}}$ for $i \ge 2$. Substituting (11) into (10) and after some manipulations, we observe that finding ${\widehat{\bf{h}}^{\left( k \right)}}$ amounts to finding
\begin{equation}
\widehat{\bf{h}}_i^{\left( k \right)} = \mathop {\arg \max }\limits_{{{\bf{h}}_i}} f\left( {\widehat{\bf{h}}_{1:i - 1}^{\left( k \right)},{{\bf{h}}_i},\widehat{\bf{h}}_{i + 1:L}^{\left( k \right)}} \right), \forall i
\end{equation}
where
\begin{equation}
\begin{array}{l}
 f\left( {\widehat{\bf{h}}_{1:i - 1}^{\left( k \right)},{{\bf{h}}_i},\widehat{\bf{h}}_{i + 1:L}^{\left( k \right)}} \right) \\
  = \mathop {\max }\limits_{{{\bf{h}}_{1:i - 1}},{{\bf{h}}_{i + 1:L}}} \left\{ {\prod\limits_{j = 1}^L {p\left( {{{\bf{h}}_j}\left| {{{\bf{h}}_{j - 1}}} \right.} \right)\prod\limits_{j = 1}^L {{e^{{Q_j}\left( {{{\bf{h}}_j}\left| {{{\widehat{\bf{h}}}^{\left( {k - 1} \right)}}} \right.} \right)}}} } } \right\} \\
  = \underbrace {\mathop {\max }\limits_{{{\bf{h}}_{1:i - 1}}} \left\{ {p\left( {{\bf{h}}_1 \left| {{\bf{h}}_0 } \right.} \right)\prod\limits_{j = 1}^{i-1} {\left( {{e^{{Q_{j }}\left( {{{\bf{h}}_{j}}\left| {{{\widehat{\bf{h}}}^{\left( {k - 1} \right)}}} \right.} \right)}}p\left( {{{\bf{h}}_j}\left| {{{\bf{h}}_{j-1}}} \right.} \right)} \right)} } \right\}}_{\left( 1 \right)}
  \times \\ \underbrace {{e^{{Q_i}\left( {{{\bf{h}}_i}\left| {{{\widehat{\bf{h}}}^{\left( {k - 1} \right)}}} \right.} \right)}}}_{\left( 2 \right)} \times \underbrace {\mathop {\max }\limits_{{{\bf{h}}_{i + 1:L}}} \left\{ {\prod\limits_{j = i+1}^{L} {\left( {{e^{{Q_j}\left( {{{\bf{h}}_j}\left| {{{\widehat{\bf{h}}}^{\left( {k - 1} \right)}}} \right.} \right)}}p\left( {{{\bf{h}}_j}\left| {{{\bf{h}}_{j - 1}}} \right.} \right)} \right)} } \right\}}_{\left( 3 \right)} \\
 \end{array}
 \end{equation}
We can now solve the M-step by a message passing algorithm that implements the max-product rule \cite{kschischang2001factor, loeliger2007factor} on a factor graph. We construct the factor graph corresponding to (10), and treat ${e^{{Q_i}\left( {{{\bf{h}}_i}\left| {{{\widehat{\bf{h}}}^{\left( {k - 1} \right)}}} \right.} \right)}}$ (used in (13)) as the input message to the variable node of ${{{\bf{h}}_i}}$. The message passing algorithm is a bidirectional algorithm consisting of forward and  backward message passing. For each direction, the message passing mechanism is very similar to Viterbi algorithm (VA) for convolutional codes except that the variables $\left\{ {{{\bf{h}}_i}} \right\}$ are continuous.  As indicated in (13), there are three messages needed for the computation of  $\widehat{\bf{h}}_i^{\left( k \right)}$: message (1) is the result of message passing in the forward direction; message (2) is the input message to ${{\bf{h}}_i}$; and message (3) is the result of message passing in the backward direction. At first sight, it may seem that to obtain the messages in (13) (in particular, in the computation associated with the max operation), we need to search over the continuous space of the variables. This would be highly complex. It turns out that that is not the case, as explained below.
\begin{figure*}[!t]
\normalsize
\includegraphics[width=7in]{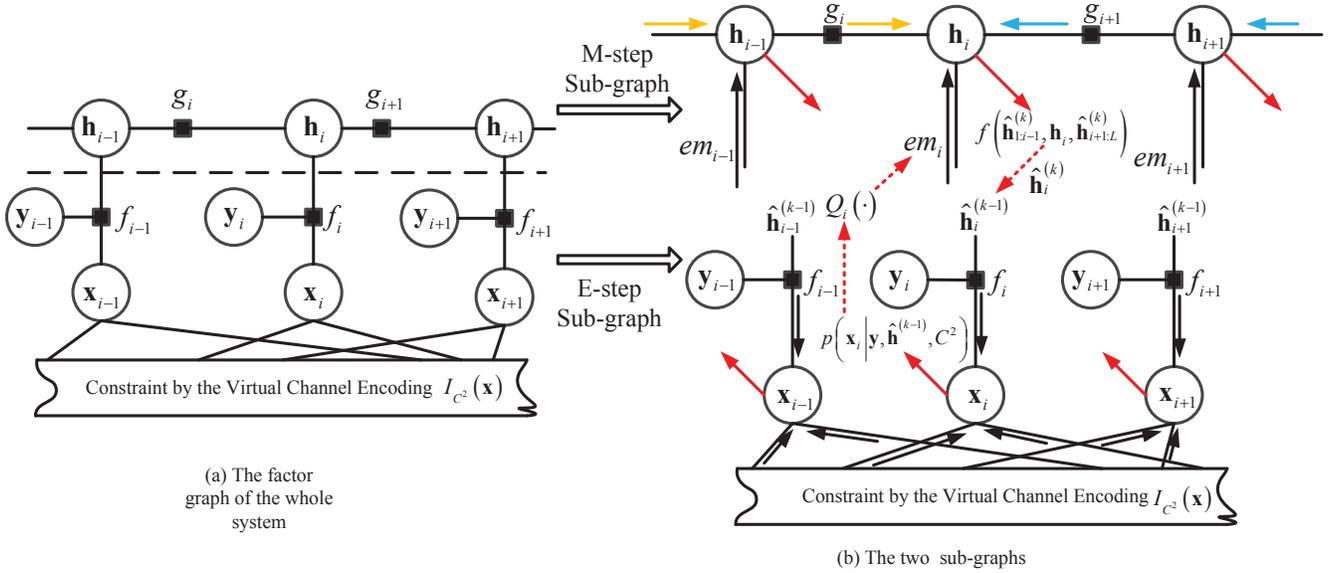}
\caption{The illustration for the implementation of EM-BP PNC on factor graph: (a) The factor graph for the whole system; (b) The two sub-graphs.} \label{fg}
\end{figure*}

Borrowing the jargon from \cite{dauwels2009expectation}, the term ${e^{{Q_i}\left( {{{\bf{h}}_i}\left| {{{\widehat{\bf{h}}}^{\left( {k - 1} \right)}}} \right.} \right)}}$ is the $i^{th}$ EM message and we abbreviate it as
$$e{m_i} = {e^{{Q_i}\left( {{{\bf{h}}_i}\left| {{{\widehat{\bf{h}}}^{\left( {k - 1} \right)}}} \right.} \right)}}.$$
For our application, this EM message has a Gaussian functional form, greatly facilitating message passing on the factor graph. To see the Gaussian form, we write
\begin{eqnarray*}
\begin{array}{l}
 \exp \left( {{Q_i}\left( {{{\bf{h}}_i}\left| {{{\widehat{\bf{h}}}^{\left( {k - 1} \right)}}} \right.} \right)} \right) \\
  = \exp \left( { - \sum\limits_{{{\bf{x}}_i}} {p\left( {{{\bf{x}}_i}\left| {{\bf{y}},{{\widehat{\bf{h}}}^{\left( {k - 1} \right)}},{C^2}} \right.} \right){{{{\left\| {{y_i} - {\bf{h}}_i^{\rm{T}}{{\bf{x}}_i}} \right\|}^2}} \mathord{\left/
 {\vphantom {{{{\left\| {{y_i} - {\bf{h}}_i^{\rm{T}}{{\bf{x}}_i}} \right\|}^2}} {\sigma _n^2}}} \right.
 \kern-\nulldelimiterspace} {N_{0}}}} } \right) \\
  \propto {\cal C}{\cal N}\left( {{{\bf{h}}_i}:{{\bf{m}}_{e{m_i}}},{{\bf{K}}_{e{m_i}}}} \right) \\
 \end{array}
\end{eqnarray*}
with mean vector and covariance matrix
\begin{subequations}\label{EMm}
\begin{equation}
{{\bf{m}}_{e{m_i}}} = {\left( {{{\bf{K}}_{{{\bf{x}}_i}}} + {{\bf{m}}_{{{\bf{x}}_i}}}{\bf{m}}_{{{\bf{x}}_i}}^{\rm{H}}} \right)^{ - 1}}{{\bf{m}}_{{{\bf{x}}_i}}}{y_i}
\end{equation}

\begin{equation}
{{\bf{K}}_{e{m_i}}} = {\left( {{{\bf{K}}_{{{\bf{x}}_i}}} + {{\bf{m}}_{{{\bf{x}}_i}}}{\bf{m}}_{{{\bf{x}}_i}}^{\rm{H}}} \right)^{ - 1}} N_{0}
\end{equation}
\end{subequations}
where
$$\begin{array}{l}
 {{\bf{m}}_{{{\bf{x}}_i}}} = \sum\limits_{{{\bf{x}}_i}} {p\left( {{{\bf{x}}_i}\left| {{\bf{y}},{{\widehat{\bf{h}}}^{\left( {k - 1} \right)}},{C^2}} \right.} \right)} {{\bf{x}}_i} \\
 {{\bf{K}}_{{{\bf{x}}_i}}} = \sum\limits_{{{\bf{x}}_i}} {p\left( {{{\bf{x}}_i}\left| {{\bf{y}},{{\widehat{\bf{h}}}^{\left( {k - 1} \right)}},{C^2}} \right.} \right)} \left( {{{\bf{x}}_i} - {{\bf{m}}_{{{\bf{x}}_i}}}} \right){\left( {{{\bf{x}}_i} - {{\bf{m}}_{{{\bf{x}}_i}}}} \right)^{\rm{H}}} \\
 \end{array}$$
are the mean and variance of ${{{\bf{x}}_i}}$  computed based on its APP.  Therefore, the model underlying the maximization problem (10) is a pure linear Gaussian model: all the messages on the factor graph turn out to be Gaussian functions of variables of interest.  The maximization problem associated with a linear Gaussian model such as (10) can be solved by Gaussian message passing \cite{loeliger2007factor} that implements the max-product rule. For Gaussian message passing, we only need to track the mean vectors and covariance matrices of messages, since together they fully characterize the Gaussian distributions. This avoids searching over the whole continuous feasible region, allowing practical implementation of the message passing algorithm for the M-step.

We have now presented the big picture on how to implement EM channel estimation as a message passing algorithm on a factor graph and to combine it with BP message passing for channel decoding. Before going into the detailed derivation of the update equations for the EM-BP message passing, we make two important remarks:
\begin{enumerate}
 \item We have shown above that, in channel-coded communication systems, the EM messages of the channels are Gaussian. As specified by (14), the mean and the variance of the $i^{th}$ EM message are linked to the APP of the $i^{th}$  transmitted symbol. This link connects the BP message passing for channel decoding with EM iterations.

 \item The APP of ${{\bf{x}}_i}$ is a product of three parts:
\begin{equation}
\begin{array}{l}
 p\left( {{{\bf{x}}_i}\left| {{\bf{y}},{{\widehat{\bf{h}}}^{\left( {{k} - 1} \right)}},{C^2}} \right.} \right) = A \times\\
    p\left( {{{\bf{x}}_i}\left| {{{\bf{y}}_{1:i - 1}},{{\bf{y}}_{i + 1:L}},{{\widehat{\bf{h}}}^{\left( {{k} - 1} \right)}},{C^2}} \right.} \right)p\left( {{y_i}\left| {{{\bf{x}}_i},{{\widehat{\bf{h}}}^{\left( {{k} - 1} \right)}}} \right.} \right), \\
 \end{array}
\end{equation}
where $A$ is a constant independent of ${{{\bf{x}}_i}}$, ${p\left( {{{\bf{x}}_i}\left| {{{\bf{y}}_{1:i - 1}},{{\bf{y}}_{i + 1:L}},{{\widehat{\bf{h}}}^{\left( {k - 1} \right)}},{C^2}} \right.} \right)}$ is the extrinsic information sent from the BP message passing for channel decoding, and ${p\left( {{y_i}\left| {{{\bf{x}}_i},{{\widehat{\bf{h}}}^{\left( {k - 1} \right)}}} \right.} \right)}$ is the information provided by the observation $y_i$. When the channel code used has cycles in its factor graph (e.g.,  LDPC, Turbo codes), the BP message passing algorithm for channel decoding requires multiple iterations to compute the extrinsic information ${p\left( {{{\bf{x}}_i}\left| {{{\bf{y}}_{1:i - 1}},{{\bf{y}}_{i + 1:L}},{{\widehat{\bf{h}}}^{\left( {k - 1} \right)}},{C^2}} \right.} \right)}$ for all $i$ \cite{kschischang2001factor}. Then, the APPs are updated according to (15).  Meanwhile, each EM iteration requires one set of such newly computed APPs for the transmitted symbols. Therefore, the theoretical construct of EM requires us to carry out the multiple iterations of BP channel decoding to update the extrinsic information as soon as the channel has been updated to a new estimate.

\end{enumerate}

Fig. 3 (a) presents the factor graph of the joint probability of our system, i. e. $p\left( {{\bf{y}},{\bf{x}},{\bf{h}}\left| {{C^2}} \right.} \right)$. On the factor graph, variable nodes $\left\{ {{{{y}}_i},{{\bf{x}}_i},{{\bf{h}}_i}} \right\}$ are denoted by circles; and check nodes $\left\{ {{f_i} \buildrel \Delta \over = p\left( {{{{y}}_i}\left| {{{\bf{x}}_i},{{\bf{h}}_i}} \right.} \right),{g_i} \buildrel \Delta \over = p\left( {{{\bf{h}}_i}\left| {{{\bf{h}}_{i  - 1}}} \right.} \right)} \right\}$ are denoted by solid squares. Further, according to the factorization $p\left( {{\bf{y}},{\bf{x}},{\bf{h}}\left| {{C^2}} \right.} \right) = p\left( {{\bf{y}},{\bf{x}}|{\bf{h}},{C^2}} \right)p\left( {\bf{h}} \right)$, we split this overall factor graph into two parts:

\noindent	\textbf{E-step subgraph}: The subgraph on the lower part of Fig. 3 (b) corresponds to the representation of $p\left( {{\bf{y}},{\bf{x}}|{\bf{h}},{C^2}} \right)$
through factorization (6). The E-step of EM is implemented by message passing on this subgraph, thus it is labeled as the E-step subgraph. Note that the subgraph that represents the constraint imposed by virtual channel encoding ${I_{{C^2}}}\left( {\bf{x}} \right)$ is embedded inside the E-step subgraph. BP message passing for channel decoding is operating on the subgraph of ${I_{{C^2}}}\left( {\bf{x}} \right)$.  Thus, we can regard BP as being a sub-step within the E-step also.

\noindent	\textbf{M-step sub-graph}: The subgraph on the upper part of Fig. 3 (b) corresponds to the representation of  $p\left( {\bf{h}} \right)$ through factorization (11). The M-step of EM is implemented by message passing on this subgraph, thus it is labeled as the M-step subgraph.

In the following, we derive the message update equations for EM-BP PNC. We denote the message sent by a node $x$  to a node $y$ by ${\mu _{x \to y}}\left(  \cdot  \right)$. We perform the BP message passing algorithm for \emph{virtual channel decoding} on the E-step subgraph to obtain the soft APP outputs $\left\{ {p\left( {{{\bf{x}}_i}\left| {{\bf{y}},{{\widehat{\bf{h}}}^{\left( {k - 1} \right)}},{C^2}} \right.} \right)} \right\}$. The BP message passing algorithm for virtual channel decoding can be derived directly by applying the sum-product rule on the factor graph that incorporates the constraints imposed by \emph{virtual channel encoding} \cite{liew2011physical, lu2012asynchronous} that models the simultaneous transmissions by terminal nodes. For virtual channel decoding within the $k^{th}$  EM iteration, the channel ${\bf{h}}$ from the M-step subgraph is fixed to the previous estimate ${\widehat{\bf{h}}^{\left( {k - 1} \right)}}$. Thus, the message ${\mu _{{f_i} \to {{\bf{x}}_i}}}$, which initializes the virtual decoding, is given by
\begin{equation}
\begin{array}{l}
 {\mu _{{f_i} \to {{\bf{x}}_i}}}\left( {{{\bf{x}}_i}} \right) = p\left( {{y_i}\left| {{{\bf{x}}_i}} \right.,\widehat{\bf{h}}_i^{\left( {k - 1} \right)}} \right)  \\ \;\;\;\;\;\;\;\;\;\;\;\;\;\;\;\;\;\;= {\cal C}{\cal N}\left( {{y_i}:{{\left( {\widehat{\bf{h}}_i^{\left( {k - 1} \right)}} \right)}^{\rm{T}}}{{\bf{x}}_i},N_{0}} \right) \\
 \end{array}
\end{equation}
for all $i$. The E-step of EM is completed by sending the EM messages $em_i$ to the M-step subgraph after BP channel decoding, as indicated by the red dotted arrows in Fig. 3 (b).  This is fulfilled by computing the mean vectors and covariance matrices of Gaussian EM messages as in (\ref{EMm}) based on the channel decoding outputs $\left\{ {p\left( {{{\bf{x}}_i}\left| {{\bf{y}},{{\widehat{\bf{h}}}^{\left( {k - 1} \right)}},{C^2}} \right.} \right)} \right\}$.

Since the structure of the M-step subgraph is a tree, the M-step (9) is exactly implemented by a forward message passing (yellow arrows in upper sub-graph in Fig. 3 (b)) and a backward message passing (blue arrows in upper sub-graph in Fig. 3 (b)). This observation also coincides with the mathematical expression for message passing by equation (13). Due to the assumed Gauss-Markov channel model and the Gaussian feature of the incoming EM messages, all the messages on the M-step subgraph preserve Gaussianity. For forward message passing, it is required to compute the message
$${\mu _{{g_i} \to {{\bf{h}}_i}}}\left( {{{\bf{h}}_i}} \right) = \mathop {\max }\limits_{{{\bf{h}}_{1:i - 1}}} \left\{ {p\left( {{{\bf{h}}_1}} \right)\prod\nolimits_{j = 2}^i {   {em_{j - 1}}    p\left( {{{\bf{h}}_j}\left| {{{\bf{h}}_{j - 1}}} \right.} \right)   }  } \right\}$$
from ${\mu _{{g_{i - 1}} \to {{\bf{h}}_{i - 1}}}}\left( {{{\bf{h}}_{i - 1}}} \right)$ recursively. Given that ${\mu _{{g_{i - 1}} \to {{\bf{h}}_{i - 1}}}}\left( {{{\bf{h}}_{i - 1}}} \right) \propto {\cal C}{\cal N}\left( {{{\bf{h}}_{i - 1}}:{\bf{m}}_{{{\bf{h}}_{i - 1}}}^f,{\bf{K}}_{{{\bf{h}}_i} - 1}^f} \right)$ and the Markovian property, we have the following update equations for forward message passing:
$$\begin{array}{l}
 {\mu _{{g_i} \to {{\bf{h}}_i}}}\left( {{{\bf{h}}_i}} \right) = \mathop {\max }\limits_{{{\bf{h}}_{i - 1}}} \left\{ {{\mu _{{g_{i - 1}} \to {{\bf{h}}_{i - 1}}}}\left( {{{\bf{h}}_{i - 1}}} \right)   e{m_{i - 1}} p\left( {{{\bf{h}}_i}\left| {{{\bf{h}}_{i - 1}}} \right.} \right)} \right\} \\
\;\;\;\;\;\;\;\;\;\;\;\;\;\;\;\;\;\;\; \propto {\cal C}{\cal N} \left( {{{\bf{h}}_i}:{\bf{m}}_{{{\bf{h}}_i}}^f,{\bf{K}}_{{{\bf{h}}_i}}^f} \right) \\
 \end{array}$$
with
\begin{subequations}\label{forward}
\begin{equation}
\begin{array}{l}
 {\bf{m}}_{{{\bf{h}}_i}}^f = \alpha {\bf{m}}_{{{\bf{h}}_{i - 1}}}^f + \alpha {\bf{K}}_{{{\bf{h}}_{i - 1}}}^f{\left( {{{\bf{K}}_{e{m_{i - 1}}}} + {\bf{K}}_{{{\bf{h}}_{i - 1}}}^f} \right)^{ - 1}} \\
\;\;\;\;\;\;\;\;\;\;\;\;\;\;\;\;\;\;\;\;\;\;\;\;\;\;\;\;\;\;\;\;\;\;\;\;\;\;  \times \left( {{{\bf{m}}_{e{m_{i - 1}}}} - {\bf{m}}_{{{\bf{h}}_{i - 1}}}^f} \right) \\
 \end{array}
\end{equation}

\begin{equation}
\begin{array}{l}
{{\bf{K}}_{{{\bf{h}}_i}}^f} = {\alpha ^2}{\bf{K}}_{{{\bf{h}}_{i - 1}}}^ f  + \left( {1 - {\alpha ^2}} \right){\bf{Q}} - \left( {\alpha {\bf{K}}_{{{\bf{h}}_{i - 1}}}^ f} \right) \\
 \;\;\;\;\;\;\;\;\;\;\;\;\;\;\;\;\;\;\;\;\; \times {\left( {{{\bf{K}}_{e{m_{i - 1}}}} + {\bf{K}}_{{{\bf{h}}_{i - 1}}}^ f } \right)^{ - 1}}\left( {\alpha {\bf{K}}_{{{\bf{h}}_{i - 1}}}^ f} \right) \\
 \end{array}
\end{equation}
\end{subequations}
The update equations (\ref{forward}) are essentially the famous Kalman one-step prediction equations. The backward update equations that compute message ${\mu _{{g_{i + 1}} \to {{\bf{h}}_i}}}\left( {{{\bf{h}}_i}} \right) $
from  ${\mu _{{g_{i + 2}} \to {{\bf{h}}_{i + 1}}}}\left( {{{\bf{h}}_{i + 1}}} \right) \propto {\cal C}{\cal N} \left( {{{\bf{h}}_{i + 1}}:{\bf{m}}_{{{\bf{h}}_{i + 1}}}^b,{\bf{K}}_{{{\bf{h}}_{i + 1}}}^b} \right)$ are easily obtained similarly by symmetry; we therefore omit them at here.  Finally, the message flowing out of variable node ${{\bf{h}}_i}$
(the red solid arrow in the upper sub-graph in Fig. 3 (b)) is the product of all the incoming messages
\begin{eqnarray*}
\begin{array}{l}
 {\mu _{{{\bf{h}}_i} \to out}}\left( {{{\bf{h}}_i}} \right) = f\left( {\widehat{\bf{h}}_{1:i - 1}^{\left( k \right)},{{\bf{h}}_i},\widehat{\bf{h}}_{i + 1:L}^{\left( k \right)}} \right) \\
  \;\;\;\;\;\;\;\;\;\;\;\;\;\; \;\;\;\;\;\;\: = {\mu _{{g_i} \to {{\bf{h}}_i}}}\left( {{{\bf{h}}_i}} \right) \times e{m_i} \times {\mu _{{g_{i + 1}} \to {{\bf{h}}_i}}}\left( {{{\bf{h}}_i}} \right) \\
 \;\;\;\;\;\;\;\;\;\;\;\;\;\; \;\;\;\;\;\;  \: \propto {\cal C}{\cal N}\left( {{{\bf{h}}_i}:{{\bf{m}}_{{{\bf{h}}_i}}},{{\bf{K}}_{{{\bf{h}}_i}}}} \right) \\
 \end{array}
\end{eqnarray*}
where the mean vector ${{{\bf{m}}_{{{\bf{h}}_i}}}}$ is used to update the new estimate of ${{\bf{h}}_i}$:
\begin{equation}
\begin{array}{l}
 \widehat{\bf{h}}_i^{\left( k \right)} = \mathop {\arg \max }\limits_{{{\bf{h}}_i}} {\mu _{{{\bf{h}}_i} \to out}}\left( {{{\bf{h}}_i}} \right) \\
  \;\;\;\;\;\;\;= {{\bf{m}}_{{{\bf{h}}_i}}} = {{\bf{m}}_{{\bf{h}}_i^ - }} + {{\bf{G}}_{{{\bf{h}}_i}}}\left( {{{\bf{m}}_{e{m_i}}} - {{\bf{m}}_{{\bf{h}}_i^ - }}} \right) \\
 \end{array}
\end{equation}
with
\begin{eqnarray*}
\begin{array}{l}
 {{\bf{K}}_{{\bf{h}}_i^ - }} = {\left( {{{\left( {{\bf{K}}_{{{\bf{h}}_i}}^f} \right)}^{ - 1}} + {{\left( {{\bf{K}}_{{{\bf{h}}_i}}^b} \right)}^{ - 1}}} \right)^{{\rm{ - }}1}} \\
 {{\bf{m}}_{{\bf{h}}_i^ - }} = {{\bf{K}}_{{\bf{h}}_i^ - }}\left( {{{\left( {{\bf{K}}_{{{\bf{h}}_i}}^f} \right)}^{ - 1}}{\bf{m}}_{{{\bf{h}}_i}}^f{\rm{ + }}{{\left( {{\bf{K}}_{{{\bf{h}}_i}}^b} \right)}^{ - 1}}{\bf{m}}_{{{\bf{h}}_i}}^b} \right) \\
 {{\bf{G}}_{{{\bf{h}}_i}}} = {{\bf{K}}_{{\bf{h}}_i^ - }}{\left( {{{\bf{K}}_{{\bf{h}}_i^ - }} + {{\bf{K}}_{e{m_i}}}} \right)^{ - 1}} \\
 \end{array}
\end{eqnarray*}
This completes the M-step of EM. We then iterate back to the E-step with the new estimate.

\subsection{Initialization and Termination of EM Iteration}

EM iteration needs to be bootstrapped with a good initial point; otherwise there is no guarantee that the algorithm will converge to the global maximum \cite{wu1983convergence}. For each block, we obtain the initial ${\widehat{\bf{h}}^{\left( 0 \right)}}$ by minimum mean square error (MMSE) estimation \cite{biguesh2006training} using only the pilot symbols. Then, the channel coefficients within each data block are simply set to the estimated channel coefficient of the closest pilot.

We repeat the E-step and M-step iteratively. When the number of iterations $k$ reaches a preset maximum limit $K$, we terminate the EM algorithm after obtaining the final channel estimate ${\widehat{\bf{h}}^{\left( K \right)}}$. Substituting ${\widehat{\bf{h}}^{\left( K \right)}}$ into the signal model (1) as the real channel ${\bf{h}}$, we could carry out a final round of channel decoding to obtain $p\left( {{\bf{x}}\left| {{\bf{y}},{{\widehat{\bf{h}}}^{\left( K \right)}},{C^2}} \right.} \right)$, where x is the overall codeword pair. The objective of this final channel decoding is consistent with Step 2 in Section III-A.  However, instead of finding $p\left( {{\bf{x}}\left| {{\bf{y}},{{\widehat{\bf{h}}}^{\left( K \right)}},{C^2}} \right.} \right)$  for channel decoding, we choose to modify Step 2 by still employing BP for virtual channel decoding to find  $p({{\bf{x}}_i}|{\bf{y}},{\widehat{\bf{h}}^{\left( K \right)}},{C^2})$ for each and every symbol pair ${\bf{x}}_i$. There are three reasons for this. First, we already have the virtual channel decoder at hand in the factor graph implementation of EM-BP PNC receiver. Second, for many advanced channel codes (e.g., LDPC, Turbo code), the decoding process finds $p({{\bf{x}}_i}|{\bf{y}},{\widehat{\bf{h}}^{\left( K \right)}}{C^2})$ rather than $p\left( {{\bf{x}}\left| {{\bf{y}},{{\widehat{\bf{h}}}^{\left( K \right)}},{C^2}} \right.} \right)$, because finding $p\left( {{\bf{x}}\left| {{\bf{y}},{{\widehat{\bf{h}}}^{\left( K \right)}},{C^2}} \right.} \right)$  for all possible codewords is generally a difficult computation-intensive problem. Finding $p({{\bf{x}}_i}|{\bf{y}},{\widehat{\bf{h}}^{\left( K \right)}}{C^2})$  for all $i$ also may be treated as a sort of approximation of the original objective 2. Third, even if we could find $p\left( {{\bf{x}}\left| {{\bf{y}},{{\widehat{\bf{h}}}^{\left( K \right)}},{C^2}} \right.} \right)$, it would not be easy to obtain the ML network-coded message because the ML network-coded codeword is given by
$${\widehat{\bf{x}}_R} = \widehat{{{\bf{x}}_A} \oplus {{\bf{x}}_B}} = \arg \mathop {\max }\limits_{{{\bf{x}}_R}} \sum\limits_{\scriptstyle \;\;\;\;\;\;\;\;{\bf{x}}: \hfill \atop
  \scriptstyle {{\bf{x}}_A} \oplus {{\bf{x}}_B} = {{\bf{x}}_R} \hfill} {p\left( {{\bf{x}}\left| {{\bf{y}},{{\widehat{\bf{h}}}^{\left( K \right)}},{C^2}} \right.} \right)}$$
where ${{\bf{x}}_A} \buildrel \Delta \over = \left\{ {{x_{A,i}}} \right\}$ and ${{\bf{x}}_B} \buildrel \Delta \over = \left\{ {{x_{B,i}}} \right\}$ are the codeword transmitted by nodes A and B. The complexity for the decoding of ${\widehat{\bf{x}}_R}$  is exponential in the length of codewords.  By contrast, the ML network-coded symbol is much easier to find from $p({{\bf{x}}_i}|{\bf{y}},{\widehat{\bf{h}}^{\left( K \right)}}{C^2})$.  It is given by
$$ \widehat{{x_{A,i}} \oplus {x_{B,i}}} = \arg \mathop {\max }\limits_x \sum\limits_{\scriptstyle \;\;\;\;\;\;\;\;\;{{\bf{x}}_i}: \hfill \atop
  \scriptstyle {x_{A,i}} \oplus {x_{B,i}} = x \hfill}  {p\left( {{{\bf{x}}_i}\left| {{\bf{y}},{{\widehat{\bf{h}}}^{\left( K \right)}},{C^2}} \right.} \right)}.$$

\begin{algorithm}[t]
\caption{The EM-BP Message Passing Implementation of Joint Channel Estimation and Channel Decoding in PNC Systems}
\begin{algorithmic}[1]
\\ \textbf{Initialization:}
\State compute ${\widehat{\bf{h}}^{\left( 0 \right)}}$ by MMSE estimation from pilot symbols;
\For{$i=1$ to $K$}
\\\textbf{The E-step of EM}
\State compute the inputs message to virtual decoding as \indent  (16), with the tentative channel estimate ${\widehat{\bf{h}}^{\left( k-1 \right)}}$;
\State  compute the soft output APPs, $p\left( {\left. {{{\bf{x}}_i}} \right|{\bf{y}},{{\widehat{\bf{h}}}^{\left( {k - 1} \right)}},{C^2}} \right)$, \indent from BP message passing for virtual channel \indent decoding; the number of iterations in the virtual \indent channel decoding is denoted by $N_{cd1}$;
\State compute the mean vectors and covariance matrices of \indent the EM messages as (14);
\\\textbf{The M-step of EM}
\State perform the forward message passing for EM channel  \indent  estimation as (17);
\State perform the backward message passing for EM  \indent  channel estimation similarly to (16);
\State compute the new channel estimate ${\widehat{\bf{h}}^{\left( k\right)}}$ as (18);
\EndFor
\\ \textbf{Termination:}
\State perform BP message passing for the final virtual channel decoding, with the final channel estimate ${\widehat{\bf{h}}^{\left(K\right)}}$; the number of iterations in the virtual channel decoding is denoted by $N_{cd2}$;
\State compute the network-coded information symbols as (19).
\end{algorithmic}
\end{algorithm}

To summarize, with the final channel estimate ${\widehat{\bf{h}}^{\left( K \right)}}$, we can use the BP message passing for virtual channel decoding in the factor graph to compute the final decoding results  $p({{\bf{x}}_i}|{\bf{y}},{\widehat{\bf{h}}^{\left( K \right)}}{C^2})$  and  $p\left( {{s_{A,j}},{s_{B,j}}\left| {{\bf{y}},{{\widehat{\bf{h}}}^{\left( K \right)}},{C^2}} \right.} \right)$
at the same time. Then, the network-coded source message is obtained by
\begin{equation}
\widehat{{s_{A,j}} \oplus {s_{B,j}}} = \arg \mathop {\max }\limits_s \sum\limits_{\scriptstyle {s_{A,j}},{s_{B,j}}: \hfill \atop
  \scriptstyle {s_{A,j}} \oplus {s_{B,j}} = s \hfill} {p\left( {{s_{A,j}},{s_{B,j}}\left| {{\bf{y}},{{\widehat{\bf{h}}}^{\left( K \right)}},{C^2}} \right.} \right)}
\end{equation}
for all $j$. After that, the relay channel-encodes the network-coded source message and broadcasts the channel-coded message to nodes A and B in the downlink phase. Obtaining the estimate of the network-coded message from  $p\left( {{s_{A,j}},{s_{B,j}}\left| {{\bf{y}},{C^2}} \right.} \right)$, which is in turn decoded by imagining a virtual encoder as the source of the information, is referred to as the joint CNC process in \cite{zhang2009channel, liew2011physical}. Whereas \cite{zhang2009channel, liew2011physical} assumes the channel coefficients are perfectly known, here we assume the channel coefficients are unknown and need to be estimated as part of the joint CNC process. Our EM-BP factor graph framework seamlessly bridges the channel estimation process and the joint CNC process. We summarize the EM-BP message passing implementation of joint channel estimation and channel decoding in PNC systems in Algorithm 1,  where ${N_{cd1}}$ denotes the number of iterations for the BP channel decoding within the EM iteration; and ${N_{cd2}}$ denotes the number of iterations for the final BP channel decoding after the termination of EM.

\section{The Application of SAGE-BP to PNC}
\begin{table*}[!t]
\caption{\label{tab:test}Complexity of EM/SAGE-BP PNC per iteration}
\centering
\begin{tabular}{ccccc}
\toprule
 & EM-BP PNC & SAGE-BP PNC\\
\midrule
channel estimation & $O\left( {2\left( {5N_u^3 + 4N_u^2 + 3{N_u}} \right)L} \right)$
 & $O\left( {24N_uL} \right)$  \\
computing EM messages   & $O\left( {\left( {N_u^3 + \left( {N_m^{{N_u}} + 3} \right)N_u^2 + \left( {2N_m^{{N_u}} + 1} \right){N_u} + 2N_m^{{N_u}} - 2} \right)L} \right)$  &$O\left( {{N_u}\left( {2N_m^{{N_u}} + 2{N_m} + 1} \right)L} \right)$\\
channel decoding  & $O\left( {{N_{cd1}}N_m^{{N_u}}\left( {2N_m^{{N_u}} - 1} \right)l} \right)$ & $O\left( {{N_u}{N_{cd1}}N_m^{{N_u}}\left( {2N_m^{{N_u}} - 1} \right)l} \right)$\\
\bottomrule
\end{tabular}
\end{table*}

SAGE sequentially updates a subset of parameters. Doing so essentially decomposes a higher dimension problem into several lower dimension sub-problems, since only a subset of the parameters are estimated and updated according to the EM mechanism each time. It can be shown that the sequential updates of subsets of parameters always still guarantee convergence \cite{sage1994}. In this section, we apply the theory of SAGE to the problem of joint channel estimation and channel decoding in PNC systems. We also extend the framework of EM-BP message passing over factor graph to SAGE-BP message passing. The motivation for introducing SAGE is to reduce the complexity of the channel estimation part of our framework. To our best knowledge, this is the first attempt to apply SAGE for channel estimation in PNC systems. In fact, in the context of general estimation problems, this is probably the first attempt that integrates the use of  SAGE and BP on a unified factor graph.

In our setting, it is natural to break up the problem of estimating $\bf{h}$ into two sub-problems, estimation of ${{\bf{h}}_A}$ and estimation of ${{\bf{h}}_B}$.  That is, we update the channel of node A, ${{\bf{h}}_A} \buildrel \Delta \over = \left\{ {{{{h}}_{A,i}}} \right\}$ and the channel of node B, ${{\bf{h}}_B} \buildrel \Delta \over = \left\{ {{{{h}}_{B,i}}} \right\}$ separately and alternatively. When updating one channel, we keep the other channel fixed. We formulate the  $k^{th}$ SAGE iteration as a two-stage process.

\noindent \textbf{The $1^{st}$ stage of the $k^{th}$ iteration}:  ${{\bf{h}}_B}$ is fixed to $\widehat{\bf{h}}_B^{\left( {k - 1} \right)}$; and ${{\bf{h}}_A}$ is updated by
\begin{equation}\label{sage1}
\widehat{\bf{h}}_A^{\left( k \right)} = \mathop {\arg \max }\limits_{{{\bf{h}}_A}} \left( {p\left( {{{\bf{h}}_A}} \right)\cdot\prod\nolimits_{i = 1}^L {{e^{{Q_{A,i}}\left( {{{\bf{h}}_{A,i}}\left| {\widehat{\bf{h}}_A^{\left( {k - 1} \right)},\widehat{\bf{h}}_B^{\left( {k - 1} \right)}} \right.} \right)}}} } \right)
\end{equation}
where
\begin{equation}\label{sage2}
\begin{array}{l}
 {Q_{A,i}}\left( {{{\bf{h}}_{A,i}}\left| {\widehat{\bf{h}}_A^{\left( {k - 1} \right)},\widehat{\bf{h}}_B^{\left( {k - 1} \right)}} \right.} \right) \\=
   \sum\limits_{{{\bf{x}}_i}} {\log p\left( {{{{y}}_i}\left| {{{\bf{x}}_i},{{\bf{h}}_{A,i}},\widehat{\bf{h}}_{B,i}^{\left( {k - 1} \right)}} \right.} \right)p\left( {{{\bf{x}}_i}\left| {{\bf{y}},\widehat{\bf{h}}_A^{\left( {k - 1} \right)},\widehat{\bf{h}}_B^{\left( {k - 1} \right)},{C^2}} \right.} \right)}  \\
 \end{array}
\end{equation}
is the $Q$  function of ${{\bf{h}}_{A,i}}$.

\noindent \textbf{The $2^{nd}$ stage of the $k^{th}$ iteration}: ${{\bf{h}}_A}$ is fixed to $\widehat{\bf{h}}_A^{\left( {k } \right)}$; and ${{\bf{h}}_B}$ is updated by
\begin{equation}\label{sage3}
\widehat{\bf{h}}_B^{\left( k \right)} = \mathop {\arg \max }\limits_{{{\bf{h}}_B}} \left( {p\left( {{{\bf{h}}_B}} \right)\cdot\prod\nolimits_{i = 1}^L {{e^{{Q_{B,i}}\left( {{{\bf{h}}_{B,i}}\left| {\widehat{\bf{h}}_A^{\left( k \right)},\widehat{\bf{h}}_B^{\left( {k - 1} \right)}} \right.} \right)}}} } \right)
\end{equation}
where
\begin{equation}\label{sage4}
\begin{array}{l}
 {Q_{B,i}}\left( {{{\bf{h}}_{B,i}}\left| {\widehat{\bf{h}}_A^{\left( k \right)},\widehat{\bf{h}}_B^{\left( {k - 1} \right)}} \right.} \right) \\ =
 \sum\limits_{{{\bf{x}}_i}} {\log p\left( {{{{y}}_i}\left| {{{\bf{x}}_i},\widehat{\bf{h}}_{A,i}^{\left( k \right)},{{\bf{h}}_{B,i}}} \right.} \right)p\left( {{{\bf{x}}_i}\left| {{\bf{y}},\widehat{\bf{h}}_A^{\left( k \right)},\widehat{\bf{h}}_B^{\left( {k - 1} \right)},{C^2}} \right.} \right)}  \\
 \end{array}
\end{equation}
is the $Q$  function of ${{\bf{h}}_{B,i}}$.

\noindent The computations of $Q$ functions (\ref{sage2}) and (\ref{sage4}) for all $i$ correspond to the E-step; the maximization problems (\ref{sage1}) and (\ref{sage3}) correspond to the M-step. We still employ BP channel decoding to compute the APPs used in the $Q$ functions of SAGE.  The framework of message passing on a factor graph is still applicable to SAGE-BP. We describe the key steps on how to transform the SAGE-BP computation onto a factor graph in the following.

SAGE breaks up the EM channel estimation into two sub-problems. Each sub-problem is solved by message passing on a factor subgraph. For the M-step of SAGE, the factor subgraph that solves (\ref{sage1}) ((\ref{sage3})) only consists of the channel variables of terminal node A (B): $\left\{ {{h_{A,i}}} \right\}$ ($\left\{ {{h_{B,i}}} \right\}$). The message passing algorithms for computing $\widehat{\bf{h}}_A^{\left( k \right)}$ and $\widehat{\bf{h}}_B^{\left( k \right)}$  are initialized with the incoming EM messages $\left\{ {e{m_{A,i}} \buildrel \Delta \over = {e^{{Q_{A,i}}\left(  \cdot  \right)}}} \right\}$ and $\left\{ {e{m_{B,i}} \buildrel \Delta \over = {e^{{Q_{B,i}}\left(  \cdot  \right)}}} \right\}$, respectively. It can be proven that Gaussianity is preserved. Namely, substituting (\ref{sage2}) into $e{m_{A,i}} = {e^{{Q_{A,i}}\left(  \cdot  \right)}}$ leads to the following Gaussian expression:
\begin{equation}\label{sageem}
\begin{array}{l}
 e{m_{A,i}} = \exp \left( {{Q_{A,i}}\left( {{h_{A,i}}\left| {\widehat{\bf{h}}_A^{\left( {k - 1} \right)},\widehat {\bf{h}}_B^{\left( {k - 1} \right)}} \right.} \right)} \right)   \\ \;\;\;\;\;\;\;\;\;\: \propto {\cal CN}\left( {{h_{A,i}}:{m_{x_{A,i}^ * }}{y_i} - {m_{x_{A,i}^ * \cdot{x_{B,i}}}}\widehat h_{B,i}^{\left( {k - 1} \right)},{N_0}} \right) \\
 \end{array}
\end{equation}
where $x_{A,i}^ * $ is the conjugate of $x_{A,i}$, and ${{m_{x_{A,i}^ * }}}$, ${{m_{x_{A,i}^ * \cdot{x_{B,i}}}}}$ are given by (25) shown at the top of the next page.
\newcounter{MYtempeqncnt}
\begin{figure*}[!t]
\normalsize
\setcounter{MYtempeqncnt}{\value{equation}}
\setcounter{equation}{24}
\begin{equation}
 \begin{array}{l}
 {m_{x_{A,i}^*}} = \sum\limits_{{x_{A,i}}} {\left( {\sum\limits_{{x_{B,i}}} {p\left( {{{\bf{x}}_i}\left| {{\bf{y}},\widehat{\bf{h}}_A^{\left( {k - 1} \right)},\widehat{\bf{h}}_B^{\left( {k - 1} \right)},{C^2}} \right.} \right)} } \right)x_{A,i}^*} {\rm{ = }}\sum\limits_{{x_{A,i}}} {p\left( {{x_{A,i}}\left| {{\bf{y}},\widehat {\bf{h}}_A^{\left( {k - 1} \right)},\widehat{\bf{h}}_B^{\left( {k - 1} \right)},{C^2}} \right.} \right)x_{A,i}^*}  \\
 {m_{x_{A,i}^*\cdot{x_{B,i}}}} = \sum\limits_{x_{A,i}^*\cdot{x_{B,i}}} {\left( {\sum\limits_{{{\bf{x}}_i}:\:x_{A,i}^* \cdot {x_{B,i}} = x} {p\left( {{{\bf{x}}_i}\left| {{\bf{y}},\widehat {\bf{h}}_A^{\left( {k - 1} \right)},\widehat{\bf{h}}_B^{\left( {k - 1} \right)},{C^2}} \right.} \right)} } \right)} x_{A,i}^* \cdot {x_{B,i}} \\
  \;\;\;\;\;\;\;\;\;\;\;\;\;\;\;= \sum\limits_{x_{A,i}^*\cdot{x_{B,i}}} {p\left( {x_{A,i}^* \cdot {x_{B,i}}\left| {{\bf{y}},\widehat{\bf{h}}_A^{\left( {k - 1} \right)},\widehat {\bf{h}}_B^{\left( {k - 1} \right)},{C^2}} \right.} \right)} x_{A,i}^* \cdot {x_{B,i}} \\
 \end{array}
\end{equation}
\setcounter{MYtempeqncnt}{\value{equation}}
\hrulefill
\vspace*{4pt}
\end{figure*}
Based on this Gaussian expression for ${\left\{ {e{m_{A,i}}} \right\}}$ and the Gauss-Markov channel model of ${\left\{ {{h_{A,i}}} \right\}}$ in (2), we can solve the M-step of the first stage (\ref{sage1}) by a Gaussian message passing algorithm that is almost similar to the one derived in Section III for EM, with the difference that the state space here only includes the channel of one terminal node. Thus, all the length-2 channel vectors, in the algorithm, are reduced to scalars. Similarly, we can derive the Gaussian expression for ${\left\{ {e{m_{B,i}}} \right\}}$ and solve the M-step of the second stage (\ref{sage3}) by the same Gaussian message passing algorithm. We round off the discussion of  SAGE-BP PNC with the following remarks:

\begin{enumerate}

\item	In (24), $p\left( {{x_{A,i}}\left| {{\bf{y}},\widehat{\bf{h}}_A^{\left( {k - 1} \right)},\widehat{\bf{h}}_B^{\left( {k - 1} \right)},{C^2}} \right.} \right)$  is the decoding result for the transmitted symbol of node A, ${{x_{A,i}}}$;  ${p\left( {x_{A,i}^* \cdot {x_{B,i}}\left| {{\bf{y}},\widehat{\bf{h}}_A^{\left( {k - 1} \right)},\widehat {\bf{h}}_B^{\left( {k - 1} \right)},{C^2}} \right.} \right)}$ can be regarded as the decoding result for the variable $x_{A,i}^* \cdot {x_{B,i}}$.  Based on these decoding results, the posterior means of ${x_{A,i}^*}$ and $x_{A,i}^* \cdot {x_{B,i}}$ are computed according to (25) and used in (24).  The intuitive interpretation of the expression for the mean of the Gaussian expression in (24) is that it corresponds to an interference cancelation process for the target channel.

\item  Comparing the EM messages in (14) and (24),  the message passing implementation of EM-BP needs to compute the mean vector and the covariance matrix of the transmitted symbol pair; that of SAGE-BP needs to compute the mean of the symbol transmitted from the target channel and the cross-correlation between the two symbols ${{x_{A,i}} ,{x_{B,i}}}$.

\item We compare the complexities of PNC in the following. Let us just focus on the algorithms within the EM/SAGE-BP iteration loop, since the complexities of initialization and termination are the same for EM-BP and SAGE-BP. The results are summarized in Table I. We denote the number of elements in the channel vector ${{\bf{h}}_i}$ by $N_u$. For our TWRC system, $N_u=2$. We denote the size of the modulation by $N_m$ (e. g., $N_m=2$ for BPSK and $N_m=4$ for QPSK). For EM-BP, the computation of (17) needs $O\left( {5N_u^3 + 4N_u^2 + 3{N_u}} \right)$ operations.\footnote{Strictly speaking, the complexities of matrix computations are different for different computing algorithm. Here, we assume that the inversion of an ${n}\times {n}$ matrix requires $O\left( {n^3} \right)$ computations. The product of an $n\times m$ matrix and an $m \times p$ matrix needs $O\left( {nmp} \right)$ computations for our analysis. Some fast algorithms of matrix computations can reduce these complexities to some extent.} Therefore, the complexity of channel estimation in EM-BP is $O\left( {2\left( {5N_u^3 + 4N_u^2 + 3{N_u}} \right)L} \right)$, where the factor $2$ is due to one forward and one backward message passing, $L$ is the frame length. By counting the operations involved in (14), we can figure out that the complexity of computing the EM messages for EM-BP is $O\left( {\left( {N_u^3 + \left( {N_m^{{N_u}} + 3} \right)N_u^2 + \left( {2N_m^{{N_u}} + 1} \right){N_u} + 2N_m^{{N_u}}} \right.} \right.$ $-\left. {\left. 2 \right)L} \right)$.  For SAGE-BP, there is no matrix operation, and the computation of (17) becomes $O\left( {12} \right)$. Since we perform once bidirectional message passing for each node, the complexity of channel estimation in SAGE-BP is $O\left( {24N_uL} \right)$. The complexity of computing EM messages for SAGE-BP as in (24) is $O\left( {{N_u}\left( {2N_m^{{N_u}} + 2{N_m} + 1} \right)L} \right)$.  From the above results, we can see that SAGE-BP simplifies the complexity of channel estimation by removing the need for matrix inversions and multiplications. Now, let us look at the channel decoding part. EM-BP updates $\widehat{\bf{h}}_A^{\left( k \right)}$ and $\widehat{\bf{h}}_B^{\left( k \right)}$ simultaneously; and requires one virtual channel decoding in each iteration. By contrast, SAGE-BP requires two virtual channel decodings, one before the update of $\widehat{\bf{h}}_A^{\left( k \right)}$ and one before the update of $\widehat{\bf{h}}_B^{\left( k \right)}$. That is, each time one of the channels is updated, the APPs of ${\bf{x}}_i$ will need to be re-computed before the other channel is updated. To analyze the complexity of virtual channel decoding, let us consider the regular repeat accumulate (RA) code \cite{divsalar1998coding} (used in our simulations). The codeword length is $l$, and there is no operation of channel decoding on the $L-l$ pilots symbols in each frame. The BP message passing requires $O\left( {N_m^{{N_u}}\left( {2N_m^{{N_u}} - 1} \right)} \right)$ computations per check node. There are $l$ check nodes on the factor graph of the RA code.  We perform ${{N_{cd1}}}$ iterations for each round of channel decoding. It follows that the complexity of channel decoding in EM-BP is $O\left( {{N_{cd1}}N_m^{{N_u}}\left( {2N_m^{{N_u}} - 1} \right)l} \right)$; the complexity of channel decoding in SAGE-BP is $O\left( {{N_u}{N_{cd1}}N_m^{{N_u}}\left( {2N_m^{{N_u}} - 1} \right)l} \right)$, where the factor $N_u=2$ is due to the two virtual channel decodings in each SAGE iteration. Therefore, SAGE-BP simplifies channel estimation but adds complexity to channel decoding. One simple method to maintain the same complexity in the channel decodings of SAGE-BP and EM-BP is to make the ${N_{cd1}}$ in SAGE-BP equal to half of the ${N_{cd1}}$ in EM-BP. This will cause SAGE-BP to suffer some performance loss. We will study this by simulations in the next section.

\item	The decoding results $p\left( {{x_{A,i}}\left| {{\bf{y}},\widehat{\bf{h}}_A^{\left( {k - 1} \right)},\widehat{\bf{h}}_B^{\left( {k - 1} \right)},{C^2}} \right.} \right)$, and ${p\left( {x_{A,i}^* \cdot {x_{B,i}}\left| {{\bf{y}},\widehat {\bf{h}}_A^{\left( {k - 1} \right)},\widehat {\bf{h}}_B^{\left( {k - 1} \right)},{C^2}} \right.} \right)}$ are both obtained based on the outputs of BP message passing for virtual channel decoding (see (25)). Another way to do this is to employ parallel interference cancelation (PIC) \cite{wang1999iterative, boutros2002iterativeframework} to compute $p\left( {{x_{A,i}}\left| {{\bf{y}},\widehat{\bf{h}}_A^{\left( {k - 1} \right)},\widehat{\bf{h}}_B^{\left( {k - 1} \right)},{C^2}} \right.} \right)$ ($p\left( {{x_{B,i}}\left| {{\bf{y}},\widehat{\bf{h}}_A^{\left( {k - 1} \right)},\widehat{\bf{h}}_B^{\left( {k - 1} \right)},{C^2}} \right.} \right)$) and then the joint probability $p\left( {{\bf{x}}_i \left| {{\bf{y}},\widehat{\bf{h}}_A^{\left( {k - 1} \right)} ,\widehat{\bf{h}}_B^{\left( {k - 1} \right)} ,C^2 } \right.} \right)$. Thus, we can still compute the decoding result for $x_{A,i}^* \cdot {x_{B,i}}$ from $p\left( {{\bf{x}}_i \left| {{\bf{y}},\widehat{\bf{h}}_A^{\left( {k - 1} \right)} ,\widehat{\bf{h}}_B^{\left( {k - 1} \right)} ,C^2 } \right.} \right)$. To compute these decoding results,  PIC employs two single-user channel decoders for the two terminal nodes. And it requires iterative message passing between the two single-user channel decoders, besides the iterative message passing within each of the channel decoder. Since PIC is well implemented by BP message passing \cite{boutros2002iterativeframework}, we can also incorporate it into our factor graph framework for SAGE-BP. This method is referred to as SAGE-BP PIC.  After the termination of SAGE iteration,  SAGE-BP PIC obtains ${\left\{ {p\left( {{s_{A,j}}\left| {{\bf{y}},\widehat {\bf{h}}_A^{\left( K \right)},\widehat {\bf{h}}_B^{\left( K \right)},{C^2}} \right.} \right)} \right\}}$, ${\left\{ {p\left( {{s_{B,j}}\left| {{\bf{y}},\widehat{\bf{h}}_A^{\left( K \right)},\widehat{\bf{h}}_B^{\left( K \right)},{C^2}} \right.} \right)} \right\}}$ from the final channel decoding. Then, we have, from the rule of sum-product algorithms, $p({s_{A,j}},{s_{B,j}}|{\bf{y}},\widehat {\bf{h}}_A^{\left( K \right)},\widehat {\bf{h}}_B^{\left( K \right)},{C^2})  \approx p({s_{A,j}}|{\bf{y}},\widehat {\bf{h}}_A^{\left( K \right)},\widehat {\bf{h}}_B^{\left( K \right)},{C^2}) \times p({s_{B,j}}|{\bf{y}},\widehat {\bf{h}}_A^{\left( K \right)},\widehat {\bf{h}}_B^{\left( K \right)},{C^2})$ for all $j$, from which we can perform network coding as in  (19). Since PIC employs single-user decoding whose complexity does not increase exponentially with the number of nodes $N_u$ (as virtual channel decoding does), SAGE-BP PIC can make our framework scalable with the number of nodes (if we want to extend our treatment to beyond TWRC). However, we will see in Section V that the performance of SAGE-BP PIC is not as good as SAGE-BP PNC. The reason is that there is a small cycle between node $x_{A,i}$ and node $x_{B,i}$ for each $i$ in the factor graph of PIC, and these small cycles degrade the performance of the BP algorithm. On the factor graph of virtual channel decoding, we cluster $x_{A,i}$ and $x_{B,i}$ together as one node ${\bf{x}}_{i}$. Since $x_{A,i}$ and $x_{B,i}$ now become a single variable node, the edges connecting them disappear. By this clustering technique \cite{kschischang2001factor},  the small cycles between $x_{A,i}$ and $x_{B,i}$  are removed from the factor graph of virtual channel decoding.

\item  We comment here that if nodes A and B employ different channel encoders, a corresponding virtual channel decoding does not exist. Then, we can only apply a channel decoding method for MUD systems (such as PIC) to compute the APP of $x_{A,i}$ and the APP of $x_{B,i}$, respectively. By replacing the virtual channel decoding with PIC channel decoding, we can deal with the set-up of different terminal nodes using different channel encoders under the EM/SAGE-BP framework.

\item  Since the complexities of both SAGE for channel estimation and BP for PIC channel decoding are scalable with the number of nodes, we can apply SAGE-BP PIC to systems where collisions of more than two signals are possible \cite{cocco2011collision, goseling2013random}. The study of this application awaits future work.

\end{enumerate}

\section{Simulation Results}

In this section, we perform computer simulations to evaluate the performances of the proposed schemes. We assume the channels of both terminal nodes have the same average power $\sigma _A^2 = \sigma _B^2$. Unless stated otherwise, the channel correlation coefficient $\alpha$ is set to $0.99$. The regular RA code with coding rate ${1 \mathord{\left/ {\vphantom {1 3}} \right.\kern-\nulldelimiterspace} 3}$ is employed.  In the case of BPSK modulation, each frame has 1024 information bits (thus, 3072 coded modulated data symbols); and in the case of QPSK modulation, each frame has 2048 bits (also 3072 coded modulated data symbols). We insert two pilots every $\Delta$ data symbols. Unless stated otherwise, the pilot interval $\Delta$ is set to 16  (this corresponds to a $11.1\%$ pilot load).  The two terminal nodes adopt orthogonal pilots, wherein ${P_1} = 1,{P_2} = 1$ for node A and ${P_1} = 1,{P_2} = - 1$  for node B. All simulation results  presented here are obtained by averaging over ${10^5}$ pairs of frames. The signal to noise ratio (SNR) is defined as ${{{E_s}} \mathord{\left/
 {\vphantom {{{E_s}} {N_{0}}}} \right.
 \kern-\nulldelimiterspace} {N_{0}}}$ where ${{E_s}}$ is the energy per coded bit. Specifically, for coding rate ${1 \mathord{\left/ {\vphantom {1 3}} \right.\kern-\nulldelimiterspace} 3}$, ${E_s} = {{{E_b}} \mathord{\left/
 {\vphantom {{{E_b}} 3}} \right.
 \kern-\nulldelimiterspace} 3}$ where ${{E_b}}$ is energy per source bit.

\subsection{Performance of EM-BP PNC}

First, we investigate the performance of the EM-BP PNC receiver. We evaluate the BER of the network-coded messages and the mean square error (MSE) of the estimated channels. The results of the PNC receiver using just a one-shot MMSE channel estimation (this is equivalent to our EM-BP PNC receiver with $K = 0$) and the ideal PNC receiver with the full channel state information (CSI) will be given as benchmarks.

\begin{figure}[!t]
\centering
\includegraphics[width=3.5in]{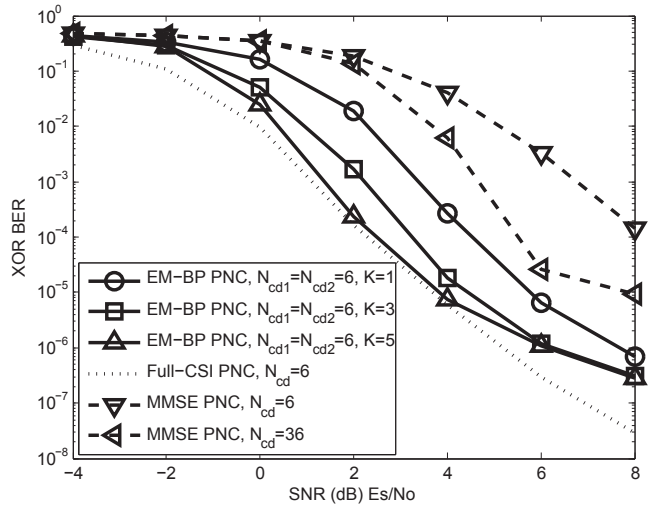}
\caption{The BER results of the EM-BP PNC receiver with BPSK modulation.} \label{simu1}
\end{figure}

\begin{figure}[!t]
\centering
\includegraphics[width=3.5in]{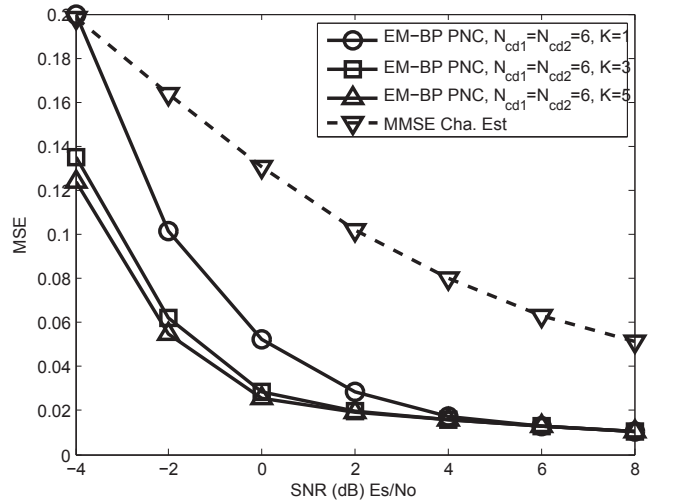}
\caption{The MSE results for the estimated channels of the EM-BP PNC receiver with BPSK modulation.} \label{simu2}
\end{figure}

Fig. \ref{simu1} presents the BER results of the three receivers: the EM-BP PNC, the MMSE PNC, and the Full-CSI PNC; and Fig. \ref{simu2} presents their MSE results for the estimated channel. The BPSK modulation is used for all receivers. The number of channel decoding iterations in the MMSE PNC receiver and the Full-CSI PNC receiver is denoted by ${N_{cd}}$. Recall that for the EM-BP PNC, $K$ is the number of EM iterations, $N_{cd1}$ is the number of iterations for the virtual channel decoding in the E-step of each EM iteration; and $N_{cd2}$ is the number of iterations in the final virtual channel decoding at the conclusion of all EM iterations. Since we use the MMSE channel estimation to initialize the EM-BP algorithm, we expect that the EM-BP algorithm to give more accurate channel estimation than the one-shot MMSE channel estimation. This is confirmed by our simulation results in Fig. \ref{simu2}. In particular, the channel estimation accuracy in EM-BP improves progressively with the number of iterations. We can also observe that the first EM iteration can already extract most of the gain in MSE. Our simulations also indicates that the EM-BP PNC algorithm has almost converged after $K=5$ iterations. These MSE improvements by EM-BP PNC are reflected into BER results. Comparing the BER results of EM-BP PNC receiver and MMSE PNC with ${N_{cd}} = 6$  in Fig. \ref{simu1}, we can see that there is a 4 dB gain by EM-BP PNC just after the first EM iteration ($K = 1$). There is a 6 dB gain after EM has converged ($K = 5$). Furthermore, the BER result of EM-BP PNC at $K = 5$ can approach the BER of the Full-CSI PNC very well.

For a fairer comparison, let us examine the performance of  EM-BP PNC  with ${N_{cd1}} = {N_{cd2}} = 6$, $K = 5$, and the performance of MMSE PNC receiver with ${N_{cd}} = 36$: i.e., the total numbers of channel decoding iterations are the same in the two cases.  We observe that for BER, EM-BP PNC receiver has around 4 dB gain over MMSE PNC. The observed error floor in both BER and MSE as SNR increases are due to the time-varying property of the channel, which is also analyzed and reported in \cite{zhu2009message}.  Essentially, even if the receiver noise is zero, the channel randomness in between pilots induces uncertainty that cannot be removed entirely regardless of the SNR. Effectively, the channel randomness in between pilots is a source of noise besides the thermal circuit noise in the receiver.

\begin{figure}[!t]
\centering
\includegraphics[width=3.5in]{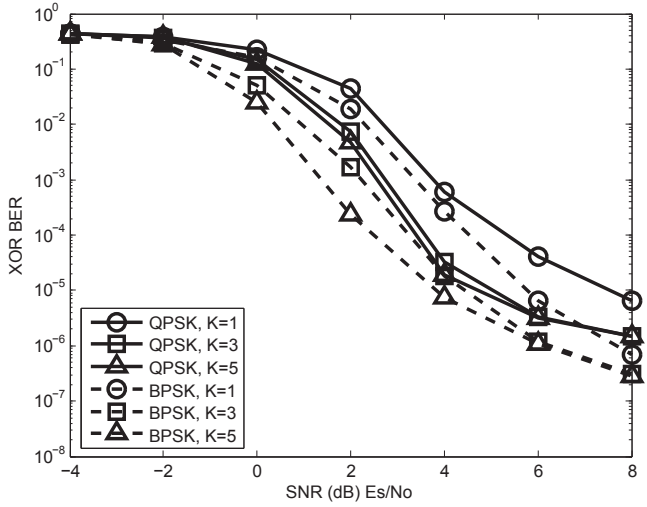}
\caption{The BER results of the EM-BP PNC receiver with QPSK modulation.} \label{simu3}
\end{figure}

\begin{figure}[!t]
\centering
\includegraphics[width=3.5in]{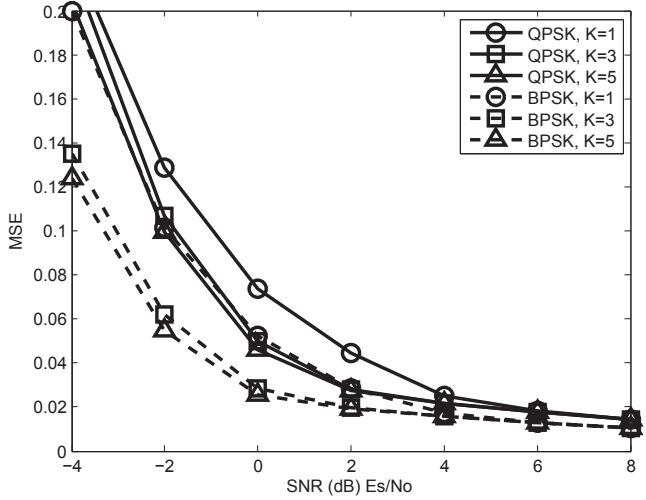}
\caption{The MSE results for the estimated channels of the EM-BP PNC receiver with QPSK modulation.} \label{simu4}
\end{figure}

Fig. \ref{simu3} and Fig. \ref{simu4} present the BER and MSE results of  EM-BP PNC when QPSK modulation is used. The results of  EM-BP PNC with BPSK are also shown as benchmarks. For the data here, ${N_{cd1}} = {N_{cd2}} = 6$. We can observe that the BER and MSE results for QPSK are slightly worse than that for BPSK. The denser constellation map in QPSK makes the channel estimation tougher. Thus, the MSEs for QPSK are larger than for BPSK. Moreover, the denser constellation makes the channel decoding more sensitive to the channel estimation error. The performance gap in the BER results of the BPSK and QPSK is about 0.5-1 dB in the low SNR regime (${{{E_s}} \mathord{\left/
 {\vphantom {{{E_s}} {{N_0}}}} \right.
 \kern-\nulldelimiterspace} {{N_0}}}<$ 6 bB), and about 2 dB in the high SNR regime (${{{E_s}} \mathord{\left/
 {\vphantom {{{E_s}} {{N_0}}}} \right.
 \kern-\nulldelimiterspace} {{N_0}}}>$ 6 bB). Since the performance trends of BPSK and QPSK are the same, we just focus on the performance of BPSK hereinafter for simplicity.

\begin{figure}[!t]
\centering
\includegraphics[width=3.5in]{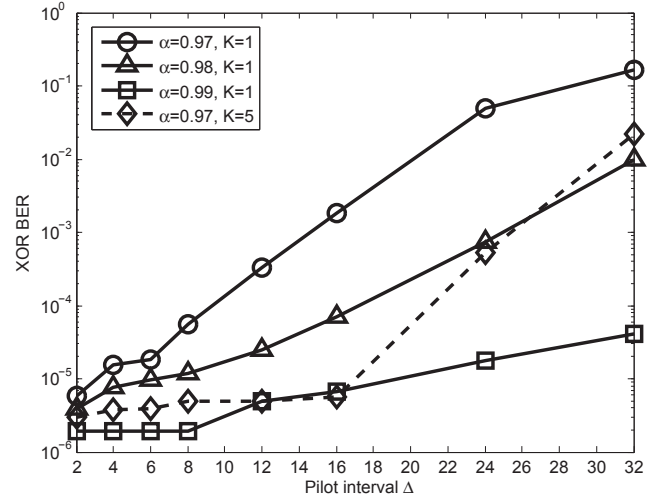}
\caption{The impact of the channel correlation coefficient $\alpha$ on the BER performance of the EM-BP PNC receiver.} \label{simu5}
\end{figure}

We now investigate the impact of channel correlation coefficient $\alpha$ on the BER performance of EM-BP PNC.  For virtual channel decoding, we also set ${N_{cd1}} = {N_{cd2}} = 6$.  The BER results of EM-BP PNC versus pilot interval $\Delta$ for various $\alpha$ are shown in Fig. \ref{simu5}. The operating SNR ${{{E_s}} \mathord{\left/
 {\vphantom {{{E_s}} {{N_0}}}} \right.
 \kern-\nulldelimiterspace} {{N_0}}}$ is 6 dB. As $\alpha$ decreases from $0.99$ to $0.97$, the channel varies faster and the BER gets worse. The BER is more sensitive to the pilot load as $\alpha$ gets smaller. When the pilot interval $\Delta$  goes from $2$ to $32$,  the BER in a channel with $\alpha=0.99$  ranges from ${10^{ - 6}}$ to ${10^{ - 5}}$; however,  the BER in a channel with $\alpha=0.97$  ranges from ${10^{ - 6}}$ to ${10^{ - 1}}$. To maintain the performance in an environment of rapidly varying channel, we can insert more pilot symbols or increase the number of EM iterations ($K$). For example, the BER of $\alpha=0.97$ and $K=5$ can approach the BER of $\alpha=0.99$ and $K=1$ when $\Delta  \le 16$.

\begin{figure}[!t]
\centering
\includegraphics[width=3.5in]{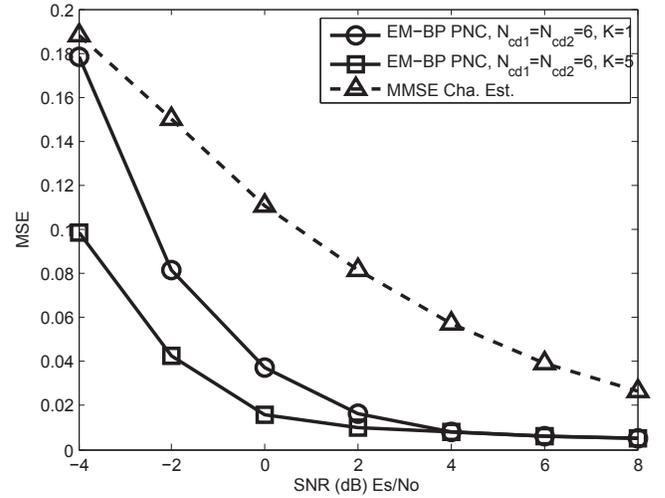}
\caption{The MSE results for the estimated channels of the EM-BP PNC receiver under the Clarke's channel (with normalized maximum Doppler spread 0.005).} \label{simu6}
\end{figure}

In Section III, we derived the proposed message passing algorithm for channel estimation using the first-order Gauss-Markov channel model. Indeed, the Gauss-Markov channel model is an approximation of the real physical channel. Here, we conduct a simulation using a more realistic mobile channel. Our aim is to demonstrate the robustness of the channel-estimation message passing when there is a mismatch between the channel model and the actual realized physical channel. All the set-ups of the simulation are the same as the previous cases except that the actual channel gains are generated according to the Clarke's channel model \cite{goldsmith2005wireless}. The normalized maximum Doppler spread in the Clarke's channel model is set to $0.005$. Following the relation between the Doppler spread and the correlation coefficient $\alpha$ established in \cite{medard2000effect}, we set the correlation coefficient to $\alpha=0.989$ and use it for channel-estimation message passing. We evaluate the MSEs of the channel estimation outputs, and present the results in Fig. \ref{simu6}. We can see that despite the channel mismatch, our channel estimation message passing still works well. We can still observe its MSE gains over the one-shot MMSE channel estimation. Comparing the MSE results of the Charke's channel in Fig. \ref{simu6} with the MSE results of the Gauss-Markov channel in Fig. \ref{simu2}, we can see that there is no obvious performance difference between them. For the Clarke's channel, besides the MSE improvement, the EM-BP PNC receiver can also yield BER improvement over the one-shot MMSE channel estimation method. We omit the BER results here to conserve space.

\subsection{Performance of SAGE-BP PNC}

\begin{figure}[!t]
\centering
\includegraphics[width=3.5in]{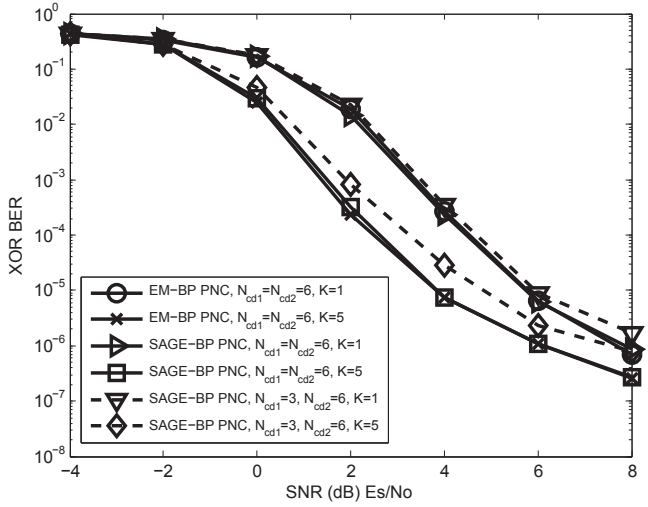}
\caption{The BER results of the SAGE-BP PNC receiver.} \label{simu7}
\end{figure}

We now investigate the performance of the SAGE-BP PNC receiver. We first focus on SAGE-BP PNC  with ${N_{cd1}} = {N_{cd2}} = 6$. The BER results are shown in Fig. \ref{simu7}, where the BER results of EM-BP PNC with ${N_{cd1}} = {N_{cd2}} = 6$ are also presented as benchmarks. It can be observed that SAGE-BP PNC has the same performance as EM-BP PNC. Both SAGE-BP PNC and EM-BP PNC converge to the same performance after $K=5$ iterations. This is consistent with the fundamental theory of SAGE \cite{sage1994}. Substituting the simulation parameters (${N_{cd1}}=6$, $N_u=2$, $N_m=2$, $l = {{8L} \mathord{\left/{\vphantom {{8L} 9}} \right.\kern-\nulldelimiterspace} 9}$) into Table I, we can find that per EM-BP iteration needs $O\left( {333L} \right)$ computations; per SAGE-BP iteration needs $O\left( {369L} \right)$ computations. The complexity of SAGE-BP now is slightly higher than that of EM-BP due to the more complex channel decoding in SAGE-BP.

For SAGE-BP PNC, since the two-dimension estimation problem of ${\bf{h}} = \left\{ {{{\bf{h}}_A},{{\bf{h}}_B}} \right\}$ in the M-step of EM has been decomposed into two one-dimension estimation sub-problems of ${{{\bf{h}}_A}}$ and ${{{\bf{h}}_B}}$, the complexity of its channel estimation is smaller than that of EM-BP PNC. To `equalize' the complexities of the channel decoding of SAGE-BP PNC and EM-BP PNC,  we set ${N_{cd1}} = 3$, ${N_{cd2}} = 6$ for SAGE-BP PNC, i.e.,  the total number of iterations for BP channel decoding in the E-step after both ${{{\bf{h}}_A}}$ and ${{{\bf{h}}_B}}$ are updated is $2\times{N_{cd1}} = 6$; and ${N_{cd1}} = 6$, ${N_{cd2}} = 6$ for EM-BP PNC. Recall that the complexity of channel estimation in SAGE-BP PNC is smaller than that in EM-BP PNC (see the complexities in Table I). Since we equalize the complexities of the channel decoding processes of EM-BP and SAGE-BP (by halving the number of channel decoding iterations in each SAGE-BP iteration), the overall complexity of SAGE-BP PNC (per SAGE-BP iteration needs $O\left( {219L} \right)$ computations) is now smaller than that of EM-BP PNC because of its less complex channel estimation. From the results in Fig. \ref{simu7}, we can see that the performance of SAGE-BP PNC  with ${N_{cd1}} = 3$, ${N_{cd2}} = 6$ is not as good as that of EM-BP PNC with ${N_{cd1}} = 6$, ${N_{cd2}} = 6$.

\begin{figure}[!t]
\centering
\includegraphics[width=3.5in]{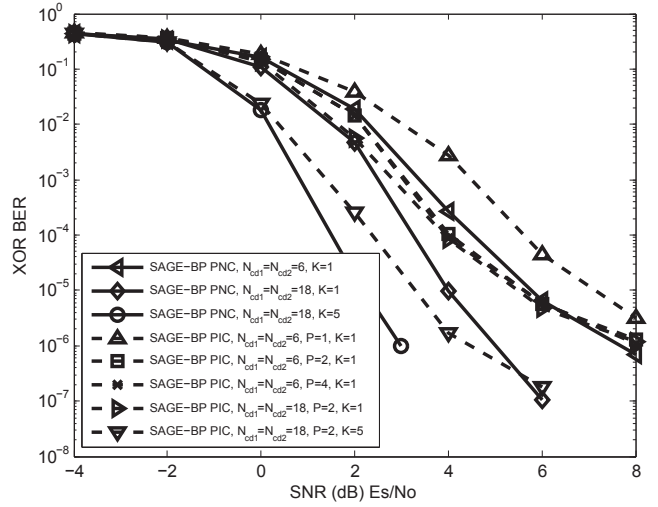}
\caption{The BER results of the SAGE-BP PIC/PNC receiver.} \label{simu8}
\end{figure}

\begin{figure}[!t]
\centering
\includegraphics[width=3.5in]{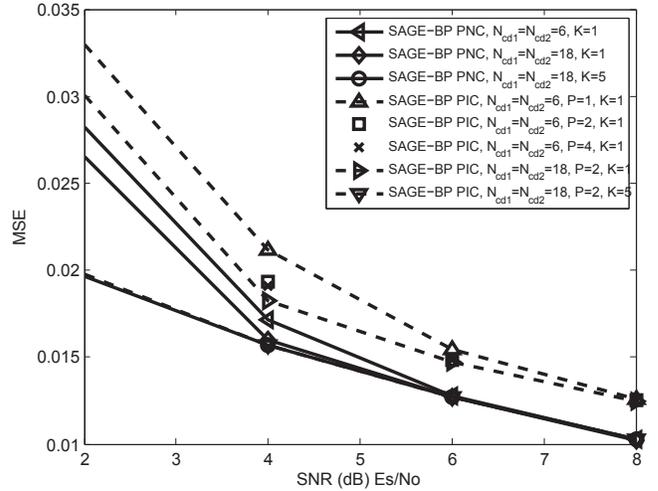}
\caption{The MSE results of the SAGE-BP PIC/PNC receiver.} \label{simu9}
\end{figure}

We next compare the performances of SAGE-BP PIC and SAGE-BP PNC. We report the BER and MSE results  in Fig. \ref{simu8} and Fig. \ref{simu9}, respectively. We denote the number of iterations for the message passing between the two single-user channel decoders in PIC by $P$.  For SAGE-BP PIC, we can increase $P$ and ${N_{cd1}}$ (${N_{cd2}}$) to allow convergence of the PIC channel decoding. From our simulation results, PIC converges with ${N_{cd1}}={N_{cd2}}=18$, $P=2$. When we continue to increase these numbers of iterations, no observable improvement on performance can be obtained.  For SAGE-BP PNC, we can increase ${N_{cd1}}$ (${N_{cd2}}$) to make the virtual channel decoding converge. Virtual channel decoding converges with ${N_{cd1}}={N_{cd2}}=18$.  From the BER results in Fig. \ref{simu8}, we can see that the BER of SAGE-BP PIC is not as good as that of SAGE-BP PNC. Specifically, the error floor of SAGE-BP PIC in the high SNR regime is higher than that of SAGE-BP PNC.  As explained earlier, this is because there are small cycles in the factor graph of PIC.  The APPs of transmitted symbols computed by PIC is not as accurate as the ones computed by virtual channel decoding for PNC.  Therefore, we expect the SAGE channel estimation based on the worse APPs will result in worse estimate results. From the MSE results in Fig. \ref{simu9}, we observe that with one SAGE iteration ($K=1$), SAGE-BP PIC indeed has worse MSE than SAGE-BP PNC. When we increase the number of iterations to allow SAGE-BP PNC and SAGE-BP PIC to converge (${N_{cd1}}={N_{cd2}}=18$, $P=2$, $K=5$), SAGE-BP PIC can obtain the same MSE as SAGE-BP PNC. However, even with the same MSE, the final channel decoding results of SAGE-BP PIC now is still worse than that of SAGE-BP PNC (see the BERs in Fig. \ref{simu8}). This is because the APPs computed by the PIC process are still not as accurate as those computed by the virtual channel decoding process. In other words, as far as the channel estimation is concerned, the performances of both schemes are comparable, but the virtual channel decoding in SAGE-BP PNC gives better estimates of the network-coded symbols.

\subsection{Comparison with Other Receiver Architectures}

In this subsection, we compare our EM-BP PNC receiver with other receiver architectures \cite{herzet2007theoretical, guo2011based, lehmann2013joint, kopbayashi2001successive}.

We first investigate the performances of two different strategies on how to combine EM with BP channel decoding: our EM-BP strategy and the strategy suggested in \cite{herzet2007theoretical, guo2011based}. Although \cite{herzet2007theoretical, guo2011based} do not investigate PNC systems, we can extrapolate their strategy for PNC application. Before we compare the performance results of the two strategies, let us explain a subtle point in the rigorous EM-BP framework because this is where the strategy in \cite{herzet2007theoretical, guo2011based} deviates from this framework.  We note that the extrinsic information used in (15) is a function of the channel estimate ${\widehat{\bf{h}}^{\left( {{k} } \right)}}$. According to the theoretical framework of EM, upon a new channel estimate, we should immediately run BP channel decoding to compute new extrinsic information to update the APPs.  The new APPs are then used to compute the next channel estimate. Our EM-BP strategy conforms to the above operational sequence. In the following discussion,  when we refer to the strategy in \cite{herzet2007theoretical, guo2011based}, we mean the strategy as applied to PNC.  In the strategy of \cite{herzet2007theoretical, guo2011based}, each iteration of BP channel decoding is followed by several EM iterations for channel estimation using the same extrinsic information from that single iteration of BP channel decoding.  For example, at EM iteration $k = {k_{1}}$, ${\widehat{\bf{h}}^{\left( {{k_{1}} - 1} \right)}}$ was obtained.  After one iteration of BP channel decoding, we obtain the new extrinsic information $p\left( {{{\bf{x}}_i}\left| {{{\bf{y}}_{1:i - 1}},{{\bf{y}}_{i + 1:L}},{{\widehat{\bf{h}}}^{\left( {{k_1} - 1} \right)}},{C^2}} \right.} \right)$  for all $i$. For several subsequent EM iterations for channel estimation, indexed by $k = {k_1} + 1, \cdots ,{k_2}-1 $, BP channel decoding will not be performed at all. EM updates a new channel estimate ${{{\widehat{\bf{h}}}^{\left( {k } \right)}}}$ in each iteration using an approximate APP of ${{{\bf{x}}_i}}$ rather than (14). Specifically,
   $$
\begin{array}{l}
 p\left( {{{\bf{x}}_i}\left| {{\bf{y}},{{\widehat{\bf{h}}}^{\left( {k - 1} \right)}},{C^2}} \right.} \right)  \approx  \\
 A \cdot p\left( {{{\bf{x}}_i}\left| {{{\bf{y}}_{1:i - 1}},{{\bf{y}}_{i + 1:L}},{{\widehat{\bf{h}}}^{\left( {{k_1} - 1} \right)}},{C^2}} \right.} \right)p\left( {{y_i}\left| {{{\bf{x}}_i},{{\widehat{\bf{h}}}^{\left( {k - 1} \right)}}} \right.} \right). \\
 \end{array}
    $$
    In particular, an approximation is made on the extrinsic information:
    $$
\begin{array}{l}
 p\left( {{{\bf{x}}_i}\left| {{{\bf{y}}_{1:i - 1}},{{\bf{y}}_{i + 1:L}},{{\widehat{\bf{h}}}^{\left( {k - 1} \right)}},{C^2}} \right.} \right)
   \\\approx p\left( {{{\bf{x}}_i}\left| {{{\bf{y}}_{1:i - 1}},{{\bf{y}}_{i + 1:L}},{{\widehat{\bf{h}}}^{\left( {{k_1} - 1} \right)}},{C^2}} \right.} \right).\\
\end{array}
    $$
Obviously, this operation dose not correctly compute the APP of ${{{\bf{x}}_i}}$ needed by EM for iterations $k = {k_1} + 1, \cdots ,{k_2}-1 $. Moreover, if there are cycles in the factor graph of the channel code (e. g. the RA code used in our simulation experiments), the computation of $p\left( {{{\bf{x}}_i}\left| {{{\bf{y}}_{1:i - 1}},{{\bf{y}}_{i + 1:L}},{{\widehat{\bf{h}}}^{\left( {{k_1} - 1} \right)}},{C^2}} \right.} \right)$  with just one BP channel decoding iteration will not accurate either. This strategy is referred to as SP-EM in \cite{herzet2007theoretical}, where SP stands for sum-product. To avoid confusion, we rename it as multiple-EM-single-BP scheme here to represent the fact that there are multiple iterations of EM for each iteration of BP. Our EM-BP is single-EM-multiple-BP in that sense.

\begin{figure}[!t]
\centering
\includegraphics[width=3.5in]{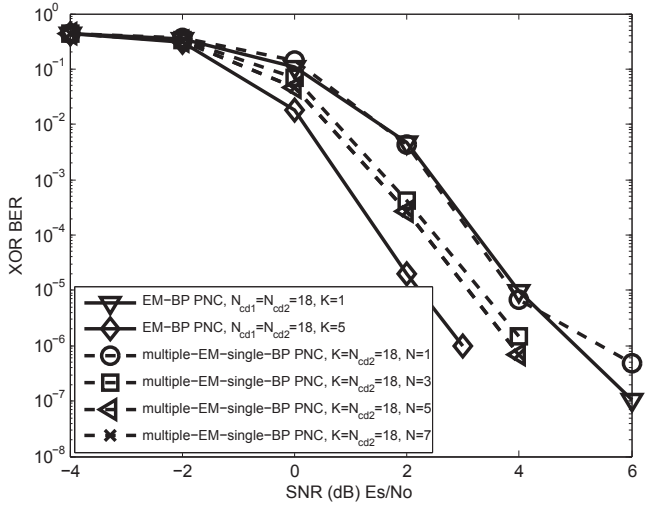}
\caption{Comparing the different strategies on how to combine EM with BP channel decoding: the BER results.} \label{simu10}
\end{figure}

\begin{figure}[!t]
\centering
\includegraphics[width=3.5in]{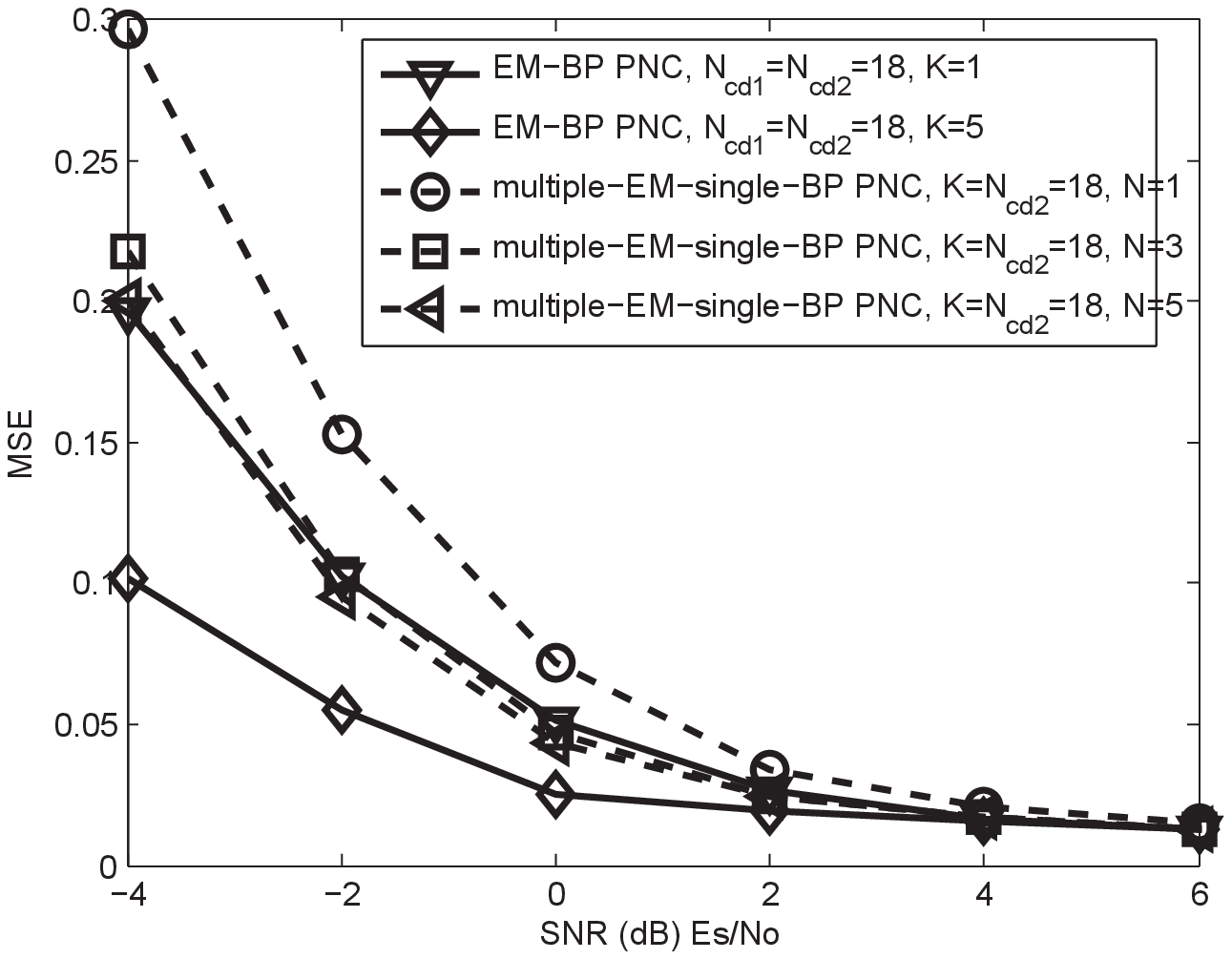}
\caption{Comparing the different strategies on how to combine EM with BP channel decoding: the MSE results} \label{simu11}
\end{figure}

The BER results of EM-BP PNC and multiple-EM-single-BP PNC are shown in Fig. \ref{simu10}; their MSE results are shown in Fig. \ref{simu11}. For EM-BP PNC, there are ${N_{cd1}}$  iterations for BP channel decoding after each EM iteration; and totally $K$ EM iterations are performed  (therefore, altogether there are $K$ EM iterations and ${N_{cd1}}K$  BP channel decoding iterations). For multiple-EM-single-BP PNC PNC, there are $K$ EM iterations after each for BP channel decoding iteration; and totally $N$ iterations for BP channel decoding are performed prior to termination (therefore, altogether there are $NK$ EM iterations and $N$ BP channel decoding iterations). After termination, we perform additional ${N_{cd2}}=18$ iterations for BP channel decoding in both strategies. The approximate APPs in multiple-EM-single-BP as explained above causes it to deviate from the principle of EM algorithm. The results in Fig. \ref{simu10} and Fig. \ref{simu11} show that compared with EM-BP PNC, multiple-EM-single-BP PNC exhibits worse performances and converges to an inferior operating point.  From our simulation results in Fig. \ref{simu10} and Fig. \ref{simu11}, we see that no further  improvement on the performance of multiple-EM-single-BP PNC can be obtained when $N$ is increased from $5$ to $7$. In other words, even if we make $N$ very large to equalize the channel decoding complexity of multiple-EM-single-BP PNC with that of EM-BP PNC, the performance of multiple-EM-single-BP PNC will still be worse than that of EM-BP PNC.

\begin{figure}[!b]
\centering
\includegraphics[width=3.5in]{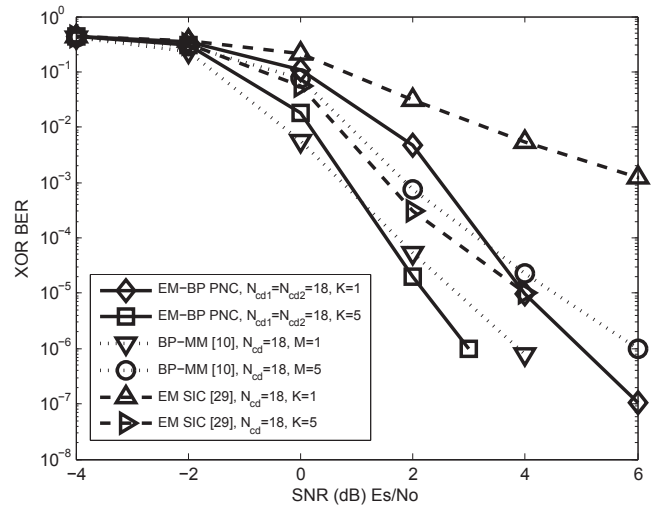}
\caption{Comparing the proposed EM-BP PNC receiver with other receiver architectures.} \label{simu12}
\end{figure}

Next, we compare our EM-BP PNC receiver and the BP-MM receiver proposed in \cite{lehmann2013joint} (see Section I for the overview of \cite{lehmann2013joint}).  In EM-BP PNC,  EM is employed to accomplish the task of channel estimation; and BP is employed to accomplish the task of channel decoding. EM can find ${\widehat{\bf{h}}_{MAP}}$ when it converges to the global optimal, and in this case the result of the final channel decoding in EM-BP PNC is $p\left( {{{\bf{x}}_i}\left| {{{\widehat{\bf{h}}}_{MAP}},{\bf{y}},{C^2}} \right.} \right)$ for all $i$.  This is the target of EM-BP. For EM, the convergence to the global optimal can always be guaranteed by a good initial point \cite{wu1983convergence}. We can also finish the tasks of both channel estimation and channel decoding using BP alone, as in \cite{lehmann2013joint}. In this case, the final target of channel decoding is  $p\left( {{{\bf{x}}_i}\left| {{\bf{y}},{C^2}} \right.} \right) = \int {p\left( {{{\bf{x}}_i},{\bf{h}}\left| {{\bf{y}},{C^2}} \right.} \right)d{\bf{h}}}$, which is different from the target of EM-BP. However, the integration over continues variables required by BP channel estimation is computationally infeasible.  BP-MM uses MM as an approximation to circumvent the need for explicit integration (i.e., integration of the approximate Gaussian distribution can be obtained in closed form) \cite{lehmann2013joint}. However, as a consequence of the approximation by MM, the optimality of BP cannot be guaranteed (i.e.,  even if the algorithm converges to the global optimal, it will not be the global optimal associated with the original non-Gaussian distribution).  The BER results of BP-MM are shown in Fig. \ref{simu12}, where $M$ denotes the number of iterations between channel estimation and channel decoding in BP-MM; and $N_{cd}$ denotes the number of channel decoding iterations in BP-MM. Comparing the BERs of EM-BP PNC and BP-MM,  EM-BP PNC is worse in the low SNR regime and it is better in the high SNR regime. Specifically, when both receivers have converged ($M=K=5$, $N_{cd}=N_{cd1}=N_{cd2}=18$), EM-BP PNC outperforms BP-MM by 1dB at the BER of $10^{-6}$. We believe this gap could be due to the approximation by MM.

As summarized in Section I, there is an EM approach for joint channel estimation and channel decoding in multi-user CDMA systems, proposed in\cite{kopbayashi2001successive}. In \cite{kopbayashi2001successive}, channel decoding is implemented by MMSE SIC with separate single-user channel decoders for data from different users; and channel estimation is implemented by EM using  APPs obtained from the single-user channel decoders. We compare EM-BP PNC with an EM SIC receiver modified from that in \cite{kopbayashi2001successive}. In the modified EM SIC receiver, there is no despreading operation, since we focus on PNC systems rather than CDMA systems. Specifically, given the channel estimate ${{{\widehat{\bf{h}}}^{\left( {k - 1} \right)}}}$ from the last EM iteration, we directly employ MMSE SIC to compute the APP of the $i^{th}$ symbol transmitted by a individual node: $p\left( {{x_{A,i}}\left| {{\bf{y}},{{\widehat{\bf{h}}}^{\left( {k - 1} \right)}}} \right.},C \right)$,  $p\left( {{x_{B,i}}\left| {{\bf{y}},{{\widehat{\bf{h}}}^{\left( {k - 1} \right)}}} \right.},C \right)$, for all $i$. MMSE SIC \cite{boutros2002iterativeframework} is a technique in which the MMSE estimation on the signal of one node is subtracted from the received signal; then channel decoding is performed on the remaining signal using the standard sum-product algorithm to obtain the APP of the other node. Based on these APPs, the channel estimation is implemented by the EM message passing as proposed in our paper here. After the final round of channel decoding, we obtain $\left\{ {{{\widehat{s}}_{A,j}}} \right\}$, $\left\{ {{{\widehat{s}}_{B,j}}} \right\}$ from the two single-user channel decoders. We then perform network coding as $\left\{ {{{\widehat{s}}_{A,j}} \oplus {{\widehat{s}}_{B,j}}} \right\}$. Compared with EM-BP PNC, there are two drawbacks to this EM SIC receiver: 1) the APPs $p\left( {{x_{A,i}}\left| {{\bf{y}},{{\widehat{\bf{h}}}^{\left( {k - 1} \right)}}} \right.},C \right)$,  $p\left( {{x_{B,i}}\left| {{\bf{y}},{{\widehat{\bf{h}}}^{\left( {k - 1} \right)}}} \right.},C \right)$, are only approximately computed by MMSE SIC (even without considering the cycles in the factor graph of channel coding) because of the use of the aforementioned MMSE signal cancelation rather than the use of the strict sum-product formalism to link the computations of $p\left( {{x_{A,i}}\left| {{\bf{y}},{{\widehat{\bf{h}}}^{\left( {k - 1} \right)}}} \right.},C \right)$ and $p\left( {{x_{B,i}}\left| {{\bf{y}},{{\widehat{\bf{h}}}^{\left( {k - 1} \right)}}} \right.},C \right)$ together; and 2) its single-user channel decoding is not optimal for PNC. The BER results of  EM SIC are shown in Fig. \ref{simu12}. when both receivers have converged ($K=5$, $N_{cd}=N_{cd1}=N_{cd2}=18$), we see that EM-BP PNC outperforms EM SIC by around 2dB at BER $10^{-5}$.

\section{Conclusion}

We have proposed an EM-BP factor-graph framework tailored for solving the problem of joint channel estimation and channel decoding in PNC systems.  This framework consists of EM message passing for channel estimation and BP message passing for channel decoding. The output of one forms the input of the other, and vice versa, so that the results of channel estimation and channel decoding can be refined in an iterative and progressive manner. A salient feature of our framework is the use of `virtual channel decoding' to ensure optimal performance for PNC systems. Furthermore, we show that the EM messages in our factor-graph framework are Gaussian messages that can be characterized by their means and variances only, and this greatly reduces computation complexity. We refer to the receiver based on this framework as EM-BP PNC.

Our simulation results indicate that the BER of EM-BP PNC can approach that of an ideal PNC receiver with perfect CSI. In addition, EM-BP PNC outperforms other receivers in terms of BER and MSE. Beyond PNC, we believe the EM-BP factor graph framework proposed in this work can also be used to construct receivers with superior performance in conventional  single-user and multi-user systems.

\appendix
\section{An interpretation of the EM algorithm}
In this appendix, we use the argument of K-L divergence to interpret the physical meaning of EM algorithm. Specifically, we shall see that the iteration expressed by (5) will eventually at least converge to a local optimum and possibly to a global optimum with respect to the target $\arg {\max _{\bf{h}}}\log p\left( {{\bf{h}}\left| {\bf{y}} \right.,{C^2}} \right)$. The objective function being optimized in the EM algorithm can be interpreted as one in which an additional Kullback-Leibler (KL) divergence term, $- {D_{KL}}$, has been added to the original objective $\log p\left( {{\bf{h}}\left| {\bf{y}} \right.,{C^2}} \right)$. This additional  $- {D_{KL}}$ term is guaranteed to converge to zero in the EM algorithm, hence the two objectives are consistent.

\noindent  \textbf{Proposition 1}: Define a function of two variables ${\bf{h}}$ and ${\bf{h}}'$ as follows: $f\left( {{\bf{h}},{{\bf{h}}'}} \right) \buildrel \Delta \over =  - {D_{KL}}\left( {\left. {p\left( {{\bf{x}}\left| {{\bf{y}},{{\bf{h}}'},{C^2}} \right.} \right)} \right\|p\left( {{\bf{x}}\left| {{\bf{y}},{\bf{h}},{C^2}} \right.} \right)} \right) + \log p\left( {{\bf{h}}\left| {{\bf{y}},{C^2}} \right.} \right)$. Furthermore, let  ${{\bf{h}}^*} \buildrel \Delta \over = \arg \mathop {\max }\limits_{\bf{h}} \log p\left( {{\bf{h}}\left| {{\bf{y}},{C^2}} \right.} \right)$
and $\left( {{\bf{h}}_f^*,{\bf{h}}_f'^{*}} \right) \buildrel \Delta \over = \arg \mathop {\max }\limits_{{\bf{h}},{{\bf{h}}'}} f\left( {{\bf{h}},{{\bf{h}}'}} \right)$.  Then, $\mathop {\max }\limits_{{\bf{h}},{{\bf{h}}'}} f\left( {{\bf{h}},{{\bf{h}}'}} \right) = \mathop {\max }\limits_{\bf{h}} \log p\left( {{\bf{h}}\left| {{\bf{y}},{C^2}} \right.} \right)$ and ${\bf{h}}_f^* = {\bf{h}}_f'^{*} = {{\bf{h}}^*}$.

\noindent \textbf{Proof}: It is known that ${D_{KL}}\left( {\left. {p\left( {{\bf{x}}\left| {{\bf{y}},{{\bf{h}}^\prime },{C^2}} \right.} \right)} \right\|\left. {p\left( {{\bf{x}}\left| {{\bf{y}},{\bf{h}},{C^2}} \right.} \right)} \right)} \right.$ $\ge 0$
for any duple $({\bf{h}},{{\bf{h}}'})$ \cite{cover2012elements}.  Thus, we have
$$
\begin{array}{l}
 f\left( {{\bf{h}},{{\bf{h}}'}} \right) \le \log p\left( {{\bf{h}}\left| {{\bf{y}},{C^2}} \right.} \right) \le \mathop {\max }\limits_{\bf{h}} \log p\left( {{\bf{h}}\left| {{\bf{y}},{C^2}} \right.} \right) \\
  \;\;\;\;\;\;\;\;\;\;\;\;\;\;\;\;\;\;\;\;\;\;\;\;\;\;\;\;\;\;\;\;\;\;\;\;\;\;\;\;\;\; = \log p\left( {{{\bf{h}}^*}\left| {{\bf{y}},{C^2}} \right.} \right) \\
 \end{array}
$$
for any duple $({\bf{h}},{{\bf{h}}'})$.  In particular,
$$
f\left( {{\bf{h}}_f^*,{\bf{h}}_f'^{*}} \right) \le \log p\left( {{{\bf{h}}^*}\left| {{\bf{y}},{C^2}} \right.} \right)
$$
We also have that  ${D_{KL}}\left( {\left. {p\left( {{\bf{x}}\left| {{\bf{y}},{{\bf{h}}^*},{C^2}} \right.} \right)} \right\|p\left( {{\bf{x}}\left| {{\bf{y}},{{\bf{h}}^*},{C^2}} \right.} \right)} \right) = 0$ \cite{cover2012elements}. Thus, setting ${\bf{h}}_f^* = {\bf{h}}_f'^{*} = {{\bf{h}}^*}$ gives us
$$
\;\;\;\;\;\;\;\;\;\;\;\;\;\;\;\;\;\;\;\;\;  f\left( {{\bf{h}}_f^*,{\bf{h}}_f'^{*}} \right) = \log p\left( {{{\bf{h}}^*}\left| {{\bf{y}},{C^2}} \right.} \right) \;\;\;\;\;\;\;\;\;\;\;\;{\rm{Q}}{\rm{.E}}{\rm{.D}}{\rm{.}}
$$

With proposition 1, algorithm (5) can be interpreted as trying to find ${\bf{h}}_f^* = {\bf{h}}_f'^{*} = {{\bf{h}}^*}$ that maximize $f\left( {{\bf{h}},{{\bf{h}}'}} \right)$. We can think about the algorithm in the following way. Since we know that the two arguments ${\bf{h}}$ and ${{\bf{h}}'}$  must be equal at the optimal, we could start out with a guess of ${\bf{h}} = {{\bf{h}}'} = {{\bf{h}}^{\left( 0 \right)}}$. This gives us an initial $f\left( {{\bf{h}},{{\bf{h}}'}} \right) = f\left( {{{\bf{h}}^{\left( 0 \right)}},{{\bf{h}}^{\left( 0 \right)}}} \right) = \log p\left( {{{\bf{h}}^{\left( 0 \right)}}\left| {{\bf{y}},{C^2}} \right.} \right)$. However, this may not be the optimal $\log p\left( {{{\bf{h}}^*}\left| {{\bf{y}},{C^2}} \right.} \right)$ even though the associated KL divergence is $0$.

In the next iteration, we want to know whether we can change ${\bf{h}}$ to a different value, say ${\bf{h}} = {{\bf{h}}^{\left( 1 \right)}}$, and obtain a better $f\left( {{{\bf{h}}^{\left( 1 \right)}},{{\bf{h}}^{\left( 0 \right)}}} \right) > f\left( {{{\bf{h}}^{\left( 0 \right)}},{{\bf{h}}^{\left( 0 \right)}}} \right)$.  This is exactly what (5) attempts to do. Notice that if such $f\left( {{{\bf{h}}^{\left( 1 \right)}},{{\bf{h}}^{\left( 0 \right)}}} \right)$ can be found, then it is guaranteed that $\log p\left( {{{\bf{h}}^{\left( 1 \right)}}\left| {{\bf{y}},{C^2}} \right.} \right) \ge f\left( {{{\bf{h}}^{\left( 1 \right)}},{{\bf{h}}^{\left( 0 \right)}}} \right) > f\left( {{{\bf{h}}^{\left( 0 \right)}},{{\bf{h}}^{\left( 0 \right)}}} \right) = \log p\left( {{{\bf{h}}^{\left( 0 \right)}}\left| {{\bf{y}},{C^2}} \right.} \right)$. Therefore,  $f\left( {{{\bf{h}}^{\left( 1 \right)}},{{\bf{h}}^{\left( 1 \right)}}} \right) = \log p\left( {{{\bf{h}}^{\left( 1 \right)}}\left| {{\bf{y}},{C^2}} \right.} \right) \ge f\left( {{{\bf{h}}^{\left( 1 \right)}},{{\bf{h}}^{\left( 0 \right)}}} \right) > f\left( {{{\bf{h}}^{\left( 0 \right)}},{{\bf{h}}^{\left( 0 \right)}}} \right)$. Thus, we see that (5) is an algorithm to successively find a better ${\bf{h}} = {{\bf{h}}'} = {{\bf{h}}^{\left( k \right)}}$ for substitution into $f\left( {{\bf{h}},{{\bf{h}}'}} \right)$ until things converge. Note in particular that $f\left( {{{\bf{h}}^{\left( {k + 1} \right)}},{{\bf{h}}^{\left( {k + 1} \right)}}} \right) \ge f\left( {{{\bf{h}}^{\left( k \right)}},{{\bf{h}}^{\left( k \right)}}} \right)$ for all $k$ by similar argument as above. Since $f\left( {{{\bf{h}}^{\left( k \right)}},{{\bf{h}}^{\left( k \right)}}} \right)$ is upper bounded by $\log p\left( {{{\bf{h}}^*}\left| {{\bf{y}},{C^2}} \right.} \right)$,  it cannot increase indefinitely and convergence is guaranteed.

However, like all other `peak seeking' algorithms, the ultimate point to which EM converges may or may not be the global peak $\log p\left( {{{\bf{h}}^*}\left| {{\bf{y}},{C^2}} \right.} \right)$  if there are local optimal points. Therefore, for global optimum, EM usually requires a good initial point, which can be achieved using pilot symbols in our problem.

\ifCLASSOPTIONcaptionsoff
  \newpage
\fi

\bibliographystyle{IEEEtran}
\bibliography{symbollevelcombining}


\end{document}